# Probing Extended Recognition Sites in Zn-Metalloproteins via Quantum Chemistry and Polarizable Molecular Dynamics

**Nohad Gresh and Jean-Philip Piquemal**
**Laboratoire de Chimie Théorique, Sorbonne Université, UMR 7616 CNRS, 75005 Paris, France.**
gresh@lct.jussieu.fr, jean-philip.piquemal@sorbonne-universite.fr

**Abstract**.
Zn-metalloproteins play vital roles in numerous metabolic processes, making them high-value targets for structure-based drug design. To advance these efforts, it is useful to unravel the individual components of the intermolecular interaction energies (ΔE) that stabilize the Zn-binding cavity, both in the absence and presence of bound protein ligands. Here, we utilize quantum chemistry (QC) to decompose ΔE into distinct physical contributions. The relative magnitudes of these components vary significantly depending on the coordination number (four to six) and the chemical nature ('hard' vs. 'soft') of the Zn-coordinating ligands. These high-level QC analyses serve to calibrate and validate polarizable molecular mechanics potentials, effectively extending the accurate description of electronic effects beyond the immediate Zn-binding cavity to enlarged recognition sites and, ultimately, entire protein systems over long molecular dynamics (MD) simulation timescales. We focus here on a QC-grounded molecular mechanics procedure, denoted as SIBFA. Following a concise overview of our QC methodology used for its calibration, we present validation studies on complexes containing up to 300 atoms and discuss the prospects of applying this framework to large-scale simulations of Zn-metalloprotein–ligand complexes. Finally, the structural and energetic roles of discrete, highly polarizable water molecules are highlighted.
.

## *Introduction.*

Zn-metalloproteins exert prominent roles in numerous metabolic events[1] . This is documented in highly informative reviews regarding enzymatic as well as non-enzymatic proteins. Although this list is far from limitative, we mention cell growth and differentiation [2] and cell repair [3]. Non-enzymatic Zn-fingers play an essential role for genome stability [4], with engineered Zn-finger proteins having been considered for therapeutic applications [5-6]. When overexpressed, Zn-metalloproteins lead to several diseases. This is the case for several enzymatic [7] [8] [9] [10] and non-enzymatic proteins, notably Zn-fingers involved in cancer growth and cell proliferation[11] and in the growth of retroviruses[12] [13] [14]. Their involvement in multiple diseases is an incentive for the design of Zn-metalloprotein inhibitors, several of which are mentioned below.

The earliest Zn-metalloenzyme inhibitors are those of Carbonic Anhydrase (CA) used against glaucoma and those of the Angiotensin Converting Enzyme (ACE) used against hypertension These were reviewed by Chen et al, and Hou et al. [8] [10]. Inhibitors of HDAC (histone deacetylase) used against malignancies could also be used against inflammatory diseases [15] . Inhibitors of neprilysin[16] and aminopeptidase N have anti-nociceptive properties [17] *and Refs. therein*. Inhibitors of matrix metalloproteinases (MMPs) could be essential against auto-immune diseases, inflammation, and tumor cell invasion during metastasis [8] [10]. An emerging target is the class of bacterial Zn-metalloenzymes, responsible for acquired resistance to bacteria against antibiotics[18] . Several inhibitors were designed in the last decade against New Delhi Metalloenzyme (NDM) and Verona-Integron Metalloenzymes (VIM-2) [19] [20] [21] [22] [23] . Among novel emerging classes are inhibitors with a triazole thione entity chelating both Zn(II) cation of the dizinc binding sites of NDM an VIM2-,[24-25] and those with a bicyclic boronate [20, 21]. Another class of Zn-metalloenzymes involved in infectious and opportunistic diseases is Zn(II) phosphomannoisomerase, PMI [26] . Inhibitors of non-enzymatic Zn-fingers, such as Zn-extractors, have potential anti-viral or anti-cancer properties [13] [14] [27] [28]

Despite the large number of Zn-metalloenzyme targets, there remains a limited number of inhibitors which past phase III are effective in clinical treatments. This short list includes inhibitors of the Angiotensin Converting Enzyme against hypertension and of Carbonic Anhydrase against glaucoma. As reviewed by Hou et al. [10] , there are no such inhibitors for HDAC, MMPs, anthrax lethal factor, botulinum toxin, while some MBL inhibitors are only now starting to enter the clinics..

Early QC studies bore on Zn(II)-mediated catalytic activities of CA by semi-empirical and ab initio QC [29] [30] [31] [32] [33]. This review focuses on non-covalent interactions. Unraveling the separate energy contributions of ΔE, the Zn-polyligand intermolecular interaction energies within the recognition site, should gain insight into the underlying physics and thus guide the design of novel inhibitors. Zn(II) could adopt variable coordination numbers (4-6) and geometries [34] [35] . The preference could be modulated by a subtle interplay of the separate energy contributions and the non-additivity of polarization and charge transfer in second order, as a function of the nature of its ligands. Historically, regarding non-covalent interactions, one of the very first ab initio QC studies of Zn(II) complexes to biologically relevant ligands, including imidazole and imidazolate along with energy decomposition analyzes for radial and angular variations, was done by Demoulin and Pullman [36] . Mono- and polyligated complexes of Zn(II) with representative ligands were studied by Jacob and Giessner-Prettre with the semi-empirical MNDO method [37] . In the context of the present work, an essential asset is the availability of the Reduced Variational Space (RVS) energy decomposition procedure [38] which at the Hartree-Fock level, affords a separation of the second-order contribution into polarization and charge-transfer. This enables` to monitor their separate radial and angular dependencies, and, most importantly, their non-additivity upon passing from mono- to polyligated complexes [39]. It can also quantify the 'softness' of Zn(II) as compared to the 'hardness' of Mg(II) in terms of the magnitude of the charge-transfer of the former, and its larger dispersion contribution [40] [41]. These concepts were introduced by Parr and Pearson as reviewed in[42] Thus, soft ions, such as Zn(II) or Cd(II), having diffuse electron clouds are prone to larger dispersion interactions, and, owing to their diffuse vacant orbitals, act as better electron acceptor in charge-transfer interactions than hard ions, such as Mg(II) and Ca(II). The RVS procedure has enabled for significant refinements and validation of the SIBFA polarizable potential as discussed below. Indeed, through its design, SIBFA intends to capture the key many-body effects at play in the Zn(II) complexes, notably polarization and charge-transfer.

We mention first some of the most salient methods to handle Zn(II) complexes in large molecular complexes. They are reviewed by several authors. Several non-polarizable force-fields were developed early on to be used for MD simulations with the CHARMM and AMBER packages [43] [44-45] [46] [47] , and are being further evolved and tested [48] [49] [50] [51] . The limitations of non-polarizable potentials were addressed [52] . QM/MM approaches were developed as an alternative [53]. Along with the SIBFA and AMOEBA polarizable potentials, which will be exposed in the next section, potentials with fluctuating charges were developed [54] . Of note is a revealing survey of the determinants of metal ion selectivity in proteins, encompassing alkali, alkaline-earth, and transition metal cations[55].

## Procedure.

*QC computations.*

The Zn-ligand and ligand-ligand intermolecular interaction energies at the Hartree-Fock (HF) level, ΔE(HF), are decomposed into four separate contributions: Coulomb ($E_C$) and exchange ($E_X$) in first-order $E_1$, and polarization ($E_{pol}$) and charge-transfer ($E_{ct}$) in second-order $E_2$. This is done with the Reduced Variational Space Analysis (RVS) [38] . $E_1$ is computed with the frozen molecular orbitals (MOs) of the interacting monomers prior to, and after, re-orthogonalization of the MOs' in the complex, for $E_C$ and $E_X$ respectively. For each monomer N, $E_1$+$E_{pol}$(N) results from the energy lowering upon relaxing its occupied and vacant MO's while freezing the occupied MO's of the other monomers. $E_1$+ $E_{pol}$(N) +$E_{ct}$(N) +Sup(N) then results from the energy lowering upon relaxing its occupied and vacant MO's and the vacant MO's of the other monomers while freezing the occupied MO's of the other monomers. Sup(N) is the (artificial) energy lowering due to the extension of the basis set of monomer N on the MO's of the other monomers. Its value is removed from that of $E_{ct}$(N) at the outcome of the RVS analysis.

The contributions of correlation/dispersion are then computed as the ΔE lowering upon passing from the HF level to correlated Moller-Plesset (MP2) level [56] , or to Density Functional Theory (DFT) and DFT augmented with a dispersion contribution [57].

The RVS procedure initially coded in the Hondo [58] , then in the GAMESS packages [59], proved instrumental into evolving the SIBFA polarizable mechanics procedure on two counts. Within $E_2$(QC), it enabled to separate $E_{ct}$ from $E_{pol}$ in a meaningful, operational fashion. It further enabled to monitor their separate evolutions upon passing from dimeric to multimolecular complexes for validation, to quantify their separate non-additivities, and assess how reliably their SIBFA counterparts could account for these.

***Polarizable force-fields.***

*SIBFA (Sum of Interactions Between Fragments Ab initio computed).*

The principal force-field used in the studies of this review is SIBFA. Its initial inception was in 1984-1986 [60] [61] . $\Delta E_{tot}$(SIBFA) is a sum of five separate contributions : multipolar and short-range repulsion in first-order, polarization and charge-transfer in second-order, then dispersion and dispersion-exchange. Their formulation is given in several papers [61] [62] [63] [64] [65] [66] [67].

We outline here some of its most salient features.

$E_{MTP}$, the electrostatic multipolar contribution, is computed as a sum of multipole-multipole terms, up to quadrupoles. For a given molecule or molecular fragment, the multipoles are derived from its ab initio QC molecular wave function and are distributed on the atoms. We resorted first to the procedure developed and published by Vigné-Maeder and Claverie [68] and, subsequently, to the procedure published by Stone[69] . It was since 2003 augmented with charge-charge, charge-dipole, and charge-quadrupole penetration terms [64] .

$E_{rep}$, the short-range repulsion contribution, is a sum of bond-bond, bond-lone pair, and lone pair-lone pair terms. It was in 2011 augmented with a three-body term in the case of Zn(II) [66]

$E_{pol}$, the polarization contribution, is computed with distributed polarizabilities. These are derived from the molecular wave function of each molecule or molecular fragment, located on the centroids of the localized molecular orbitals (LMO), namely the chemical bonds and heteroatom lone-pairs, by a procedure due to Garmer and Stevens [70] and stored, along with the multipoles, in a library of fragments. The polarizing field is screened with a Gaussian function of the distance between the polarizing atom of an interacting molecule and the centroid.

$E_{ct}$, the charge-transfer contribution, was initially derived from a perturbation formula by Murrell et al. [71], as a function of the overlap between the MO's of the electron donating molecule and the virtual MO's of the acceptor molecule and of the electrostatic potential undergone by the electron donor and the electron acceptor. It also depends, in its denominator, upon the difference between the ionization potential (IP) of the donor and the electronic affinity of the acceptor, each modulated by the electrostatic potential which each undergoes in the intermolecular complex. The electron donors are the sole lone pairs, since these are the MO's with both the smallest IPs and the largest overlap with an incoming H-bond acceptor. We use the formulation published in 1995 in light of the RVS analyses of Zn(II)-ligand complexes [63].

$E_{disp}$, the dispersion correction, is formulated as a sum of $1/R^6$, $1/R^8$, and $1/R^{10}$ terms. It has contributions from the lone-pairs and embodies an explicit exchange-dispersion term. Its formulation was given by Creuzet et al. [62] and recast as Supp. Info in [72] .

Each SIBFA contribution is non-isotropic: $E_{MTP*}$ and $E_{pol}$ due to the inherent anisotropies of the multipolar potential and field, and $E_{rep}$, $E_{pol}$, $E_{ct}$, and $E_{disp}$ due to the contributions of the lone-pairs. The non-isotropy of each contribution was validated for several molecules or molecular fragments by comparing its in- and out-of-plane angular dependencies to those of its QC counterpart. By contrast, atom-atom pair interactions are inherently isotropic. In second order, in SIBFA both $E_{pol}$ and $E_{ct}$ are non-additive, as will be recalled below.
Prior to reviewing other polarizable potentials, we mention for completeness at this stage a recent paper benchmarking non-polarizable potentials, namely several variants of the Lennard-Jones or AMBER force-fields to handle Zn(II) complexes. These appeared essentially destined to properly describe Zn(II) coordination in mono-zinc and dizinc metalloprotein binding sites, but not its intermolecular interaction energies with its coordinating ligands[73]

*Comparisons with other polarizable force-fields for Zn(II) complexes.*
AMOEBA (Atomic Multipole Optimized Energetics for Biomolecular Applications) is conceptually the closest polarizable potential to SIBFA. It itself resorts to distributed ab initio multipoles and has polarizabilities distributed on the atoms instead of the LMO centroids. The polarizing field, also calculated with the distributed multipoles, is screened using a formulation due to Thole [74]. It embodies a van der Waals contribution for the sum of short-range repulsion and dispersion, resorting to a $1/R^{14}$-$1/R^{7}$ atom-atom formulation[75] [76] . It lent itself to an MD study of the hydration of divalent metal cations [77]. It subsequently enabled MD simulations and free energy calculations for the binding of inhibitors to a matrix metalloproteinase, MMP13 [78]. AMOEBA was more recently augmented with a charge-penetration model [79] similar to the 2003 SIBFA one and with a charge-transfer contribution. The latter is a pair-wise exponential function destined to fit, in dimeric complexes, the difference between the induction contribution from the SAPT2+ perturbation procedure [80] and $E_{pol}$(AMOEBA), thus devoid of the inherent anisotropy of $E_{ct}$ conferred by the lone-pair donors and of its non-additivity feature. It was validated on water clusters and by MD simulations on liquid water but not benchmarked for metal cation complexes. A novel formulation of $E_{ct}$ in AMOEBA was recently published by Wang et al.[81] as a general damping-based charge-dipole model. A calibration for several atomic anions and cations, including Zn(II), was provided. Benchmarks for a series of dimeric complexes were provided, however, not including Zn(II) at this stage, and subsequently lent itself to an MD simulation on liquid water.

In an attempt to account for charge-transfer in Zn(II) complexes, a distance-dependent reduction of the dicationic charge was introduced by Sakharov and Lim [82] . This enabled for an improved representation of Zn(II) coordination at the outcome of MD simulations on several Zn-metalloproteins. However, no benchmarking regarding intermolecular interaction energies as compared to QC was reported.

The water complexes of mono- and divalent cations encompassing Zn(II) were studied using a Drude polarizable force-field by Yu et al. [83] . Their solvation enthalpies in clusters with n=1-6 waters were reported, showing close agreements with available $E_{int}$(QC) results. Results from MD simulations for these ions in water solution lent themselves to comparisons with experimental results regarding several observables. For Zn(II) complexes, extensions to other ligands than water could be instructive.

A fluctuating charge model for Zn-containing metalloproteins was recently put forth [54] resorting to a charge-equilibration model [84]. It enabled for reliable reproduction of Zn(II) coordination at the outcome of MD simulations in a series of Zn-metalloprotein complexes. Here again, however, no benchmarking against $E_{int}$(QC) was provided.

Charge-transfer models for Zn(II) and Mg(II) in water were developed by Soniat et al. and calibrated along with a polarization contribution by energy decomposition analyses of their complexes with one water molecule [85]. No further benchmarking of $E_{int}$ as compared to $E_{int}$(QC) in polyligated complexes was provided in this case also.

## Results and discussion.

### *Monoligated Zn(II) complexes.*

Monoligated Zn(II) complexes are essential to calibrate the relevant Zn(II) parameters of the SIBFA procedure at its successive evolution stages.

The RVS procedure enabled to monitor the radial and in-plane and out-of-plane angular dependencies in the Zn(II) complexes with neutral ligands : water, formamide, imidazole, methanethiol, and anionic ones : formate, methanethiolate and hydroxy [63] . For each ligand, each ΔE(SIBFA) contribution could closely match the radial and angular dependencies of its QC counterpart. The angular dependencies could be accounted for by the distributed multipoles regarding $E_{MTP}$ vs $E_C$, and the lone pairs regarding $E_{rep}$ and $E_{ct}$. The shallow radial dependency of $E_{ct}$(SIBFA) was seen to match that of $E_{ct}$(QC). An extension to other closed-shell dications, Mg(II), Ca(II), and Cd(II), was subsequently reported [41]. The calibration of $E_{disp}$ was done for it to reproduce the MP2 vs. RVS energy difference.

The 'soft' character of Zn(II) and Cd(II) as contrasted to the 'hard' character of Mg(II) and Ca(II) [40] was conferred predominantly by the larger magnitudes of $E_{ct}$ and $E_{disp}$.

These calculations resorted to the CEP-4-31G(2d) basis set by Stevens et al.,[86] subsequently shown to reliably account for the energy trends found with more extended basis sets.

### *Polyligated Zn(II) complexes.*

*1. Dependencies of ΔE upon the Zn(II) coordination number.*

Preliminary studies on polyligated complexes were reported in the two previous papers. A subsequent, essential, validation of SIBFA in this respect was reported by Garmer et al. [87] . It bore on the Zn(II)-binding site of thermolysin (TLN) having two His- and one Glu- end side-chains (denoted as HHE) and its complexes with hydroxamate, cysteinate, and methylphosphoramidate, the Zn(II)-binding groups of several Zn-metalloenzyme inhibitors, denoted as X, S, and P respectively. Two competing coordination numbers, n, were considered: 5 vs. 6 for the Zn-HHEX and Zn-HHEP complexes and 4 vs. 5 for Zn-HHES, corresponding to mono- and bidentate binding to the formate end-side chain of the glutamate residue. At the outcome of energy-minimization, single-point QC calculations were performed at the HF and MP2 levels with the CEP 4-31G(2d) basis set. The results are reported in Table I. The SIBFA ΔE values without and with dispersion were shown to reproduce those of ΔE(HF) and ΔE(MP2), respectively, with relative errors <2%. For HHEX and HHEP, the QC preferences in favor of n=5 over than n=6 were closely accounted for by ΔE(SIBFA), as was the case for n=4 over n=5 in the HHES complex. The contrasting preferences of the summed $E_C/E_{MTP}$ and $E_X/E_{rep}$ denoted as $E_1$, versus the summed $E_{pol}$ and $E_{ct}$, denoted as $E_2$, are noteworthy, $E_2$ being in all cases the determinant in favor of the lesser coordination number. For Zn-HHEX, $E_1$ has a very small (1.5 kcal/mol out of 550) preference in favor of the lesser n value (5), but with Zn-HHES and Zn-HHEP, it is in favor of the larger n values, a preference overcome by the inverse preferences of $E_2$. Within $E_2$, it is $E_{pol}$ that dictates its preference in favor of the smaller n, as a result of the lesser screening of field exerted by Zn(II) on each ligand by the fields exerted by the other ones. $E_{ct}$ in each complex has no preference in favor of either coordination number. Increasing by one of the number of electron-donating groups upon passing from mono- to bidentate binding to formate is counteracted by the non-additivity of $E_{ct}$ owing to its dependency upon the electrostatic potential exerted on both the electron-donor lone-pairs and on the electron acceptors in the polyligated complex, as well as by increased Zn(II)-formate equilibrium distances. $E_{corr}$(MP2) and $E_{disp}$(SIBFA) display similar shallow trends, both favoring by < 1kcal/mol the larger n with HHEX, but disfavoring it by the same amount in the Zn-HHES and Zn-HHEP complexes.

**Table I**. Values of the QC and SIBFA intermolecular interaction energies in four-, five-, and six-coordinate Zn-HHEX, Zn-HHES, and Zn-HHEP model geometries. Energies in kcal/mol.

| | Zn-HHEX | | Zn-HHES | | Zn-HHEP | |
|---|---|---|---|---|---|---|
| | 5-Coordinate | 6-coordinate | 4-Coordinate | 5-Coordinate | 5-Coordinate | 6-Coordinate |
| | | | | | | |
| QC | | | | | | |
| **ΔE(HF)** | **-651.6** | **-647.0** | **-627.7** | **-624.6** | **-623.6** | **-620.1** |
| $E_{corr}$ | -54.1 | -54.9 | -65.9 | -65.4 | -72.0 | -70.9 |
| **ΔE(MP2)** | **-705.7** | **-701.9** | **-693.6** | **-690.0** | **-695.6** | **-691.0** |
| | | | | | | |
| SIBFA | | | | | | |
| $E_1$ | -555.7 | -554.2 | -512.6 | -517.9 | -529.8 | -537.1 |
| $E_{pol}$ | -73.4 | -68.8 | -88.4 | -81.0 | -80.6 | -72.3 |
| $E_{ct}$ | -17.0 | -17.2 | -22.6 | -22.7 | -16.5 | -16.0 |
| $E_2$ | -90.4 | -86.0 | -111.0 | -103.7 | -97.1 | -88.3 |
| **ΔE** | **-646.1** | **-640.2** | **-623.7** | **-621.6** | **-626.9** | **-625.4** |
| $E_{disp}$ | -50.5 | -50.9 | -59.4 | -58.7 | -53.5 | -52.5 |
| **$\Delta E_{tot}$** | **-696.6** | **-691.0** | **-683.1** | **-680.3** | **-680.4** | **-678.0** |

Extensions to polyligated complexes of Zn(II) with five and six water molecules and the formate anion in several competing arrangements were subsequently reported [88] and carried out to polyligated complexes with ‘hard’ ($H_2O$ and $OH^-$) and ‘soft’ ($CH_3SH$ and $CH_3S^-$) ligands and to two models of binuclear Zn-metalloenzymes[89].

There is another balance between the energy contributions in complexes of the Zn-metalloprotein phosphomannose isomerase (PMI) with four hydroxamate-based inhibitors. Three competing poses were considered at the outcome of molecular dynamics then energy-minimization for each complex, and validation studies done by single-point QC calculations on enlarged recognition sites having up to 265 atoms including 28 waters [90]. We subsequently focused on the sole Zn(II)-binding site [91] having nine interacting entities: Zn(II), hydroxamate (X), and the end side chains of Asn111, His113, Lys136, Glu138, His285, Tyr287, and Asp300.

For the three poses, the complexes of the hydroxamate moiety of the best-bound inhibitor, 5-phospho-D-arabinohydroxamate, 5-PAH [92], in the Zn(II)-binding site, are represented in **Supp. Info 1** from El Khoury et al. (J. Comput. Chem. 2017, 38, 1897, Figure 10).

In the first complex, A, X is bound bidentate to Zn(II), which is also bound by formamide, imidazole, and formate, namely, the end side-chains of Asn111, His113, and Glu138. In the second complex, B, X is monodentate bound, through its N-bound oxygen, to Zn(II). The C-bound O is replaced for Zn(II) coordination, by the hydroxyl O of the phenol group and His285 is replaced by His113. In the third complex, C, Zn(II) remains bound to the C-bound O of X, but no longer by the phenyl group. The coordination number is 4. The C-bound hydroxamate O has replaced the anionic O of Glu138 as a proton acceptor from Lys136, while this O atom now accepts a proton from the NH bond of X instead.

All three complexes have in common an H-bond between the end side-chain of Tyr287, the hydroxyl of the phenol ring and one anionic O of the end side-chain of Asp300. The comparisons with respect to single-point QC calculations are reported in **Table II.** The comparisons are here limited to the HF/RVS level and ΔE(SIBFA) without dispersion, to highlight again the trends for $E_1$ and $E_2$ according to the coordination number.

There is an extremely close match between ΔE(RVS) and ΔE(SIBFA), the relative error being <1%. Both procedures favor the ordering C > A > B, again conform to that of $E_{pol}$. With both QC and SIBFA, the C vs. A preference is now reinforced by $E_{ct}$ but is opposed by $E_1$.

**Table II**. Values of the QC and SIBFA intermolecular interaction energies of hydroxamate within the Zn(II) recognition site of PMI extracted from three energy-minimized complexes, A, B, and C, of PMI with a hydroxamate-based inhibitor (see text and Figure for definition). Energies in kcal/mol.

| | **A** | | **B** | | **C** | |
|---|---|---|---|---|---|---|
| | **QC** | **SIBFA** | **QC** | **SIBFA** | **QC** | **SIBFA** |
| $E_1$ | -592.2 | -583.6 | -570.3 | -575.1 | -580.5 | -578.4 |
| | | | | | | |
| $E_{pol}$(VR)/$E_{pol}$ | -102.9 | -99.6 | -100.6 | -98.7 | -110.3 | -102.5 |
| $E_{ct}$* | -27.3 | -37.3 | -39.2 | -41.0 | -36.2 | -40.8 |
| $E_2$ | -130.2 | -136.9 | -139.8 | -139.7 | -146.5 | -143.3 |
| | | | | | | |
| **ΔE** | **-723.1** | **-720.5** | **-710.1** | **-714.8** | **-727.1** | **-721.6** |

*2. Non-additivity.*

Non-additivity, $\delta E_{nadd}$, is a distinctive feature of polarizable force-fields. It could very significantly impact the preferences of ligand complexes with Zn-metalloprotein binding sites in favor of one over alternative competing arrangements. $\delta E_{nadd}$ originates predominantly from the second-order contributions, $E_{pol}$ and $E_{ct}$. While there are also contributions to $\delta E_{nadd}$ from the dispersion contribution [93] , these are not fully unraveled at this time. $\delta E_{nadd}$ under the form of a three-body short-range repulsion was also put forth for $E_{rep}$ in polyligated complexes of some cations, such as Zn(II) [66] , but its magnitude is smaller than that of $E_2$.

The strong cooperativity of $E_2$ was unraveled for multiply H-bonded complexes of water [94] [95] and of formate with formamide oligomers [96]. An essential concern related to its *anti*-cooperativity in polyligated complexes of divalent metal cations. It appears from the outset that in such complexes, the very strong polarizing field exerted by the cation on any ligand should be opposed by the summed, predominantly negative, fields exerted by the other ligands, even though these have smaller magnitudes than the field generated by Zn(II). $E_{pol}$ depends upon the square of the norm of the summed vector fields and its magnitude could be significantly affected by their presence. A proper formulation of the screening of the polarizing field is also necessary. It is done in the context of SIBFA with a Gaussian function with as ligand-specific parameters: the prefactor, the exponent, and the atom-type effective radii of the two interacting atoms or atomic centers. These are calibrated on the basis of QC results in mono-ligated complexes as reported firstly in [63] [41]. It was essential to evaluate how well could the formulation and screening calibration of $E_{pol}$ reproduce $\delta E_{nadd}$(QC) upon passing to multimolecular complexes. A less obvious issue related to $E_{ct}$, and how well including in its formulation the potentials (rather than the fields) separately undergone by the electron donors and the electron acceptors in the multimolecular complexes could reliably account for $\delta E_{nadd}(E_{ct})$ from QC calculations. This was addressed in several papers, as early as 1995. We provide below the latest validation [91]. It resorted to the aug-cc-pVTZ(-f) basis set.[97] It bore on three representative complexes in Zn-metalloproteins, with an increased summed charge on the ligands of 0, -1, and -2. With the one-letter code, they are denoted $[Zn(HHHH)]^{2+}$, $[Zn(HHE)]^{+}$, and Zn(HHCC). The first and third are found in Zn-finger proteins and the second one in the Zn(II)-binding site of the anthrax lethal factor. The first two columns of results concern $E_{pol}$(QC) and $E_{pol}$(SIBFA) and the last two ones $E_{ct}$(QC) and $E_{ct}$(SIBFA). For each complex, the first row reports the summed value of each contribution in all dimeric Zn-ligand and ligand-ligand complexes, the second row reports the actual value of the contribution and the third highlights the resulting difference. For all three complexes, the SIBFA $\delta E_{nadd}$ values of both contributions reproduce closely the trends and magnitudes of their QC counterparts. They both have lesser magnitudes upon passing from the first complex with four neutral H ligands to the second with a net negative charge of -1 (HHE), but significantly increased magnitudes upon passing to the complex with two negative charges (HHCC). The magnitude of $\delta E_{nadd}(E_{ct})$ increases faster than that of $\delta E_{nadd}(E_{pol})$ when the summed ligand charges pass from 0 to -2.

**Table III**. Non-additivity of the second-order contributions, $E_{pol}$ and $E_{ct}$ in the polyligated complexes of Zn(II) with: four His end side-chains, two His and one Glu end side-chains and two His and two Cy- end side chains. Σ denotes the summed value of the contribution in the Zn(II)-ligand and ligand-ligand complexes in the corresponding geometry. Energies in kcal/mol.

| | $E_{pol}$(RVS)/$E_{pol*}$ | | $E_{pol(}$VR)/$E_{pol}$ | | $E_{ct}$ | |
|---|---|---|---|---|---|---|
| | | | | | | |
| [Zn(HHHH)]$^{2+}$ | | | | | | |
| | | | | | | |
| | QC | SIBFA | QC | SIBFA | QC | SIBFA |
| Σ | -262.3 | -254.3 | -270.5 | -266.1 | -68.9 | -64.4 |
| | -172.2 | -176.2 | -130.5 | -131.0 | -34.0 | -25.9 |
| **δE$_{nadd}$** | **90.1** | **78.2** | **140.2** | **135.1** | **34.9** | **38.5** |
| | | | | | | |
| [Zn(HHE)]+ | | | | | | |
| | QC | SIBFA | QC | SIBFA | QC | SIBFA |
| Σ | -211.5 | -222.5 | -220.9 | -233.6 | -62.5 | -56.6 |
| | -128.8 | -136.5 | -108.4 | -109.8 | -33.6 | -26.2 |
| **δE$_{nadd}$** | **82.7** | **80.1** | **112.5** | **123.8** | **28.9** | **30.4** |
| | | | | | | |

| Zn(HHCC) | | | | | | |
|---|---|---|---|---|---|---|
| | QC | SIBFA | QC | SIBFA | QC | SIBFA |
| Σ | -315.6 | -308.8 | -325.0 | -324.4 | -135.7 | -125.5 |
| | -130.7 | -123.9 | -97.9 | -88.9 | -48.0 | -38.6 |
| **δE$_{nadd}$** | **184.8** | **185.0** | **227.1** | **235.5** | **87.7** | **86.9** |
| | | | | | | |

### *Validations on ligand-Zn-metalloprotein complexes.*

***B-lactamase of B.fragilis.***

We report the first validation results which bore on complexes of the dizinc metallo-beta-lactamase (MBL) of *B. fragilis* with the D- and L-isomers of two inhibitors, captopril and thiomandelate [98] [99] . Since 2005, we resorted to a formulation having $E_{MTP}$ augmented with charge-charge, charge-dipole, and charge-quadrupole penetration terms, thereafter denoted as $E_{MTP*}$ [64]. The inhibitors are represented in **Figure 1.** We performed SIBFA energy-minimization (EM) on a 104-residue model of MBL. The backbone was frozen and solvation was accounted for by a continuum reaction field procedure in which the electrostatic and polarization components are computed with the same distributed multipoles as for $E_{MTP}$ and $E_{pol}$ [100] [101]. Eleven starting points were considered for D- and L-thiomandelate and nine for the D- and L-isomers of captopril, each with different coordinations of the $S^-$ and of one carboxylate O to one, or to both, Zn(II) cations. At the outcome of EM, we performed in parallel single-point SIBFA and QC calculations on the complexes of the ligands with an MBL recognition site encompassing the end side-chains of the Zn(II)-coordinating residues and of Lys184, as well as the backbone and side-chain of Asn193. Representations of the complexes of thiomandelate and captopril are given in Figures 1a-c of Ref. 99 (Antony et al., 2005). The comparisons between ΔE(SIBFA) without dispersion with ΔE(QC/RVS) with the CEP-4-31G(2d) basis set and between $\Delta E_{tot}$(SIBFA) including dispersion with ΔE(QC/DFT) and ΔE(MP2) are discussed in [99]. We provide in **Figure 2** graphs of the compared evolutions of $\Delta E_{tot}$(SIBFA) and ΔE(QC/DFT) and ΔE(QC/MP2). The SIBFA curve has a near-parallelism with the MP2 and DFT curves with the same basis sets, and is encompassed between the two.

***ß-lactamase of L1.***

We next focused on complexes of the MBL of L1 bacterial strain with inhibitors having as a *dizinc* binding moiety triazole thione (TZT), a motive used in the design of a novel class of MBL inhibitors [102]. The X-ray structure of the complex of L1 MBL with a TZT-based inhibitor[103] is represented in **Figure 3**. TZT is a pentacycle with an extracyclic S atom and substituents on one of the N atoms ortho to the CS bond. Its monoanionic charge is delocalized predominantly on the S atom and on the other, non-substituted N ortho to the CS bond. **Figure 4** represents the complex of TZT complex with the dizinc site. The N atom binds to the Zn(II) atom which is complexed by three L1 His residues and the S atom binds to the Zn(II) atom which is complexed by two His and one Asp L1 residue.

In the prospect of simulations of MBL with TZT-based inhibitors, we have first calibrated TZT by probing it with two probes: a dicationic one, Zn(II), and a dipolar one, water, in a diversity of radial and in- and out-of-plane positions of approach. The TZT parameters for each SIBFA contribution were parametrized in order to match their QC counterparts: first at the HF/RVS level, regarding the SIBFA vs. their RVS counterparts, and then upon including correlation/dispersion resorting to SAPT-DFT [104] and B97-D3 functionals [105] .

Validations were done subsequently on six multimolecular complexes :

A : the complete TZT dizinc complex ; B : the unligated dizinc complex ; C : the complex of TZT with the sole mono-zinc $(His)_3$ complex ; D : the unligated $(His)_3$ site ; E : the the TZT complex with the $(Asp-(His)_2)$ complex ; and F ; the unligated $(Asp-(His)_2)$ site. We report in **Figure 5** the separate evolutions of each SIBFA contribution along with its RVS counterpart. There remained some error compensations, notably for complexes A and B regarding $E_{pol}$ and $E_{ct}$ within $E_2$. These were discussed in the paper by Kwapien et al.[106] . Nevertheless the relative error of ΔE(SIBFA) vs. ΔE(QC) (cc-pVTZ(-f) and aug-cc-pVTZ(-f)) remained < 2 %. Relative errors not exceeding 3 % were found for all six complexes upon passing to the correlated B97-D3 level, then resorting in SIBFA to correlated multipoles and polarizabilities along with the inclusion of $E_{disp}$.

In order to ensure for transferability of the TZT parameters, we considered next two other derivatives. The first has a $-CH_3$ substituent replacing the $-NH_2$ one, and the second has in addition a $-CF_3$ substituent ortho to the $-CH_3$ one. We have considered the dizinc site of VIM-2 strain (Verona Integron MBL) instead of that of L1, with a cysteinate replacing one His of the $Asp(His)_2$ site. A representation of the complexes of the three ligands is given in **Figure 6.**

Table IV lists the values of ΔE(SIBFA) and ΔE(QC) at the uncorrelated level. The three TZT derivatives are denoted as A-$NH_2$, B-$CH_3$, and C-$CF_3$. The affinity ordering of ΔE(QC/HF): B-$CH_3$ > A-$NH_2$ > C-$CF_3$ is reproduced by ΔE(SIBFA), the relative error remaining 2 %. The contrasting trends of $E_1$ and $E_2$ are noteworthy. Thus B-$CH_3$ is favoured by $E_1$, owing to the larger magnitudes of $E_C$/$E_{MTP*}$ due to the electron-donating properties of the -$CH_3$ substituent, but $E_2$ has its smallest magnitude in this complex. A-$NH_2$ has intermediate values of both $E_1$ and $E_2$. C-$CF_3$ has the largest magnitude of $E_2$ owing to its increased polarizability, but this is counteracted by its smallest $E_1$, due to the electron-withdrawing nature of the -$CF_3$ substituent.

**Table IV.** Values of the QC and SIBFA intermolecular interaction energies of three TZT Zn-binding groups in the dizinc binding site of VIM-2 Zn-metallo-enzyme. Energies in kcal/mol.

| | A-$NH_2$ | B-$CH_3$ | C-$CF_3$ |
|---|---|---|---|
| | | | |
| **ΔE(QC)** | **-1113.0** | **-1117.5** | **-1106.9** |
| **ΔE(SIBFA)** | **-1094.5** | **-1100.3** | **-1088.8** |
| $E_1$ | -795.2 | -802.4 | -782.3 |
| $E_2$ | -299.3 | -297.9 | -306.5 |

### *Accounting for solvation effects*

An early indication of the need to account for solvation effects was in a study of the relative affinities for a Zn(II) metalloprotease related to thermolysin of a thiolate and of a phosphoramidate inhibitor [107]. It had been shown experimentally that mutation of a protonated His231 residue into an Ala one abolished the binding affinity of the phosphoramidate inhibitor, but had no impact on that of the thiolate one[108] . This could only be accounted for, in the context of SIBFA simulations, if solvation effects resorting to a multipolar Continuum procedure [100 101] were accounted for. For the thiolate inhibitor, the contribution of $\Delta G_{solv}$ was shown to be able to compensate for the loss of ligand-protein ΔE in vacuo which resulted from the mutation. This compensation did not occur in the case of the phosphoramidate inhibitor, resulting into an unfavourable energy balance due to the His231→ Ala mutation. Subsequently, the need to account for 'discrete', structural waters, in addition to Continuum solvation, was put forth upon studying the complexes of the FAK kinase with five inhibitors in the pyrrolopyrimidine series [109]. Such molecules mediate through cooperativity the interactions of the inhibitors with key FAK side-chains of the recognition site. They were shown to be indispensable to properly account for the ranking of the inhibitor affinities with respect to experiment.

### 1. *Structural waters in Superoxide Dismutase.*

Superoxide Dismutase (SOD) is a Zn(II)/Cu(I) metalloenzyme which catalyzes the dismutation of $O_2$ radical into dioxygen and hydrogen peroxide [110] . Cu/Zn-SODs are essential to prevent the effect of oxidative stress in aerobic cells. A representation of the bimetallic binding site is given in **Figure 7.** X-ray structures of SOD showed the existence of highly structured waters in the vicinity of this site, extended along two chains toward the protein surface, binding to the side-chains of Thr135 and Arg141 residues, to the main carbonyls, and connecting Arg141 to Glu131 [111] . We have quantified the role of such waters in the stabilization of SOD [112]. Using a 'lighter' version of the SIBFA potential, we performed first six independent short-duration molecular dynamics (MD) simulation in the presence of additional water molecules (up to 296), with the protein backbone held rigid [*see* [112] *for details*]. At their outcome, energy-minimization was performed, resulting into complexes denoted a-f. We then extracted an extended recognition site from the energy-minimized structures a-f. Along with the two cations and their ligands, it included those residues bound by the structural waters, and up to 28 water molecules, whence a total of 301 atoms. ΔE(SIBFA) was recomputed for each, along with single-point QC at both HF and DFT-d levels with the CEP-4-31G(2d) basis set. A representation of the water network for the best bound complex c is given in in **Figure 8**. The waters with the highest dipole moments, μ, are highlighted. These values are in the range 2.75-3.01 Debye. The summed values of the dipole moments of the 28-water clusters (permanent plus induced) are given in **Figure 9** for complexes a-f. The μ values computed with SIBFA follow closely the trends of the QC μ values. **Figure 10a** shows the evolution of ΔE(SIBFA) to very closely match that of ΔE(HF), with a near-constant offset of 3%. **Figures 10b** and **10c** show the contrasting trends of $E_1$ and $E_2$ within ΔE(SIBFA). **Figure 11a** reports the corresponding evolutions of $\Delta E_{tot}$(SIBFA) and ΔE(B97-D). The two curves display here again a near-parallelism. **Figure 11b** shows a corresponding parallelism between $E_{disp}$(SIBFA) and the ΔE(QC) lowering upon passing from HF to B97-D. Finally **Figures 12a** and **12b** report the evolutions of the SIBFA and QC stabilization energies due to the 28 waters. There are very close SIBFA/QC agreements regarding such stabilization energies at both uncorrelated and correlated levels.

Such validations lend credence to the ligand-Zn metalloprotein simulations with structural, highly polarized water molecules. We will report three of these. The first two bear on phosphomannose isomerase (PMI) and the third on VIM-2 bacterial MBL.

### 2. *Complexes of PMI with D-Mannose 6-Phosphate surrogates.*

PMI is a Zn-metallo-enzyme, responsible for several diseases of microbial and parasitic origins and opportunistic infections. Its three-dimensional structure is represented in **Figure 13**. Four active site ligands of PMI having a central sugar backbone were synthetized and tested at the Laboratoire de Chimie Biorganique and Bioinorganique in Orsay, France [113] . Their structures and denominations are represented in **Figure 14.** The first three ones have a dianionic side-chain which binds at the entrance of the cavity to two cationic residues, Arg304 and Lys310. In order to increase resistance to enzymatic degradation and improve cellular permeability, a ligand having a dianionic malonate replacing the phosphate was synthesized. The four ligands are denoted **1** to *4*, respectively. **4** is endowed with a ten-fold increase of its binding affinity]. We reported in [114] the results of comparative energy balances at the outcome of SIBFA energy-minimization of the PMI-ligand complexes in presence of a Continuum solvation. The energy balances were detailed in [114]. Among the phosphate ligands, a significant preference was found in favour of **1** over **2** and **3**, consistent with the experimental results showing that whereas **1** is a high-affinity PMI substrate, neither **2** nor **3** can act as substrates. On the other hand, the malonate derivative **4** was also found to have a lesser affinity than **1,** contrary to the experimental result. This led us, in line with our previous study on FAK-inhibitor complexes [109] to include explicitly a limited number (9) of water molecules around the accessible polar sites of the complex. This was done using the 'discrete' software [115] interfaced into the SIBFA code. Their positions were re-optimized with the SIBFA potential with limited-duration MD, and EM was subsequently resumed on the ligand-PMI complexes now with the additional presence of the nine waters. The structures of the water networks at the outcome of the new EM cycles in the PMI-**4** and PMI-**1** complexes are represented in **Figures 15a** and **15b** respectively. Three networks of structured waters were found. In the case of the PMI-**4** complex, the first, with $W_1$ and $W_2$ connects one malonate oxygen to $Trp_{18}$ on the N-terminal side at the floor then to $Glu_{48}$, $Lys_{100}$, $Glu_{294}$ and $Tyr_{287}$. The second, with $W_3$, $W_4$, and $W_5$, connects another malonate O to $Asp_{17}$ at the N-terminal side. The third, with $W_7$, $W_8$, and $W_9$, connects a third malonate O to $Asp_{300}$ on the C-terminal side. Thus all three malonate-connected networks start each with a different malonate O and end with one PMI anionic O, which belongs, respectively, to $Asp_{17}$, $Glu_{48}$, and $Asp_{300}$. They span three PMI anionic residues which are distant by 18 Å, far from the actual recognition site. In the case of the PMI-**1** complex, owing to the lesser accessibility of the phosphate than of the malonate, the first network no longer starts with an anionic O, but with $Lys_{310}$ instead. Energy balances reported in[114] taking into account desolvation of the ligands with discrete waters resulted into a preferential complexation of **4** rather than **1**. The second-order contributions $E_{pol}$ and $E_{ct}$, together with $E_{disp}$, proved essential contributors to shift the balance. As in the case of FAK kinase and SOD, in the case of PMI-**4**, some discrete waters were found with very high dipole moments, six having μ values in the range 2.75-3.15 D.

The need to account for discrete waters was underlined in subsequent papers in the context of MD simulations on the SARS-CoV-2 main protease dimerization interface [116] [117-118].

*3. Complexes of PMI with four structurally related hydroxamate-containing ligands.*

In a following study [90], we considered phosphoro-D-arabinonohydroxamate, 5-PAH, a sub-micromolar inhibitor of PMI, two derivatives with one hydroxyl removed, and a third in which all three hydroxyls were removed. These are denoted as 3d-5PAH, 2d-5-PAH, and 5-PPH. 5-PAH and its three derivatives are represented in **Figure 16** and denoted a-d**.**
All three derivatives had a thousand-fold decrease in their affinities, whether with one, or with all three hydroxyls removed. It could be desirable to account for this result by theoretical computations. However, the structure of the bound PMI-5PAH complex was not known at the time of this study. We nevertheless needed to benchmark the SIBFA accuracy as compared to QC regarding the compared PMI affinities of ligands differing solely by the presence of hydroxyl substituents. We proceeded similar to our 2011 study, resorting to the same high-resolution X-ray structure of PMI [119]. The procedure of short MD followed by EM was detailed in [90]. For 5-PAH, the starting positions were from a previous study [120]. Up to 28 structural waters were included with the 'discrete' procedure and first relaxed by short MD followed by EM. A short MD on the entirety of a 164-residue model of PMI, 5-PAH, Zn(II) and the structural waters was then done, the PMI backbone being held rigid. Three poses were selected and then submitted to EM in the presence of the Langlet-Claverie Continuum procedure to complete accounting for solvation effects. A representation of the extended recognition site at the outcome of EM for the three poses is given in **Supp. Info 2** (Gresh et al. J. Comput. Chem. 2016, 37, 2770; Figures 2a-c) and are denoted A-C. Such structures significantly differ in terms of Zn(II)-coordination. Extensive water networks are seen in all complexes connecting N- and C-terminal residues as was the case with the mannose-phosphate surrogates [114] [Gresh et al., 2011]. Complexes A-C were used as starting points for EM on the PMI complexes with derivatives b-d. There is thus a total of 12 differing complexes which can differ significantly in their geometries, Zn(II) coordination, and water networks. How reliable in such a case could be ΔE(SIBFA) when benchmarked against $\Delta E$(QC) at HF then at correlated levels? Extended recognition sites totalling up to 264 atoms including the 28 structural waters were extracted and single-point SIBFA and QC calculations performed. We report in **Figure 17a** the compared evolutions, firstly of $\Delta E$(SIBFA) and $\Delta E$(HF) with CEP-4-31G(2d) and cc-pVTZ basis sets, and then of $\Delta E_{tot}$(SIBFA) and $\Delta E$(DFT-d) with two functionals, B97-D3 and B3LYP-D3.

A counterintuitive result is the fact that complexes A, with bidentate coordination of hydroxamate, are not the preferred binding modes, which, instead, are found to be complexes C. Nevertheless, at both HF and DFT-d levels, $\Delta E$(SIBFA) retained very close overlaps with the QC curves. For completeness**, Figure 17b** provides four additional curves, in which the multipoles and polarizabilities used in SIBFA are computed with an augmented basis set, namely cc-pVTZ. The first curves are for $\Delta E$(SIBFA) and $\Delta E$(HF/cc-pVTZ). The second are for $\Delta E_{tot}$(SIBFA) and $\Delta E$(DFT-d/cc-pVTZ) with B97-D3 and B3LYP functionals. The close agreement with B97-D3 is retained with the cc-pVTZ basis set. The third curves are for the energy differences with respect to the best-bound complex, C-b. The fourth curves concern single-point computations of complexes A-a till C-d when the 28 waters are removed. It is then complexes A-a till A-d that are the most stable ones, namely with bidentate coordination of hydroxamate to Zn(II). *Thus, it is the network of discrete, polarizable waters that shifts the balance in favour of complexes C-a till C-d.*

Therefore, while within the mannose series, the water networks shifted the energy balance in favour of the malonate surrogate, in the present hydroxamate series of four structurally related analogues, these networks shifted the balance in favour of structure C over structure A, which would have otherwise prevailed in their absence.

The X-ray structure of PMI bound to 5-PAHz was reported after completion of this work [121]. 5PAHz is a hydrazide analogue of 5-PAH, in which the anionic hydroxamate is replaced with a neutral polar hydrazide group. Ligand binding was shown to produce significant shifts of three PMI domains with respect to the native PMI conformation. This motivates ongoing MD studies starting with this conformation, and using the AMOEBA software[78] and the Tinker-HP package[122] [123] . We will evaluate if MD can unravel alternative binding modes to the one found by X-ray crystallography, the evolution of the water networks, and plan to perform subsequent validation by QC of $\Delta E$(AMOEBA) and $\Delta E$(SIBFA) on recognition sites extracted from the most relevant poses.

*4. Complexes of the VIM-2 MBL with five TZT ligands with two differing side-chains.*

The last validations bear on the complexes of VIM-2 with five derivatives of triazole thione, which were designed, synthesized and tested at the University of Montpellier**.** They are represented in **Figure 18**. They bear each two arms. The first, R1, substitutes the C atom ortho to the unsubstituted N and to the hydrazone end. The second, R2, substitutes the hydrazone end. R1 has a phenyl ring, unsubstituted (II), or substituted with hydroxyl and/or methoxy (I, III, V) or two Cl atoms (IV). R2 has one or two benzyl rings substituted by either two hydroxyls (III) or one anionic $–CO_2^-$ (derivatives I, II, IV and V). R1 and R2 extend beyond the dizinc binding site. R1 binds to aromatic residues $Phe_{61}$ and $Trp_{87}$ and an anionic one, $Asp_{119}$. R2 interacts with the side-chains of $Arg_{228}$ and $His_{263}$, and the main-chain atoms of $Gly_{225}$ and $Asn_{233}$. The diversity of interaction modes adding up with those occurring in the dizinc binding site constitutes a demanding test case for polarizable potentials. We resorted to the X-ray crystal structure of VIM-2 complexed by inhibitor I [124] as a starting point for EM on the five VIM-2 inhibitor complexes. Fourteen discrete water molecules were retained, based on proximity criteria from the X-ray structure. EM was done on all intermolecular variables for the ligands, the two Zn(II) ions, and the 14 waters. At their outcome, an extended recognition was extracted. It encompassed the entirety of the inhibitors, the dizinc binding site, all VIM-2 residues which interact with R1 and R2, and the fourteen waters. It totals about 300 atoms. These sites are represented in **Supp. Info 3** (Kwapien et al., J. Comput. Chem. 2021, 42, 86; Figure 2). Three different single-point QC and SIBFA calculations were performed on the five complexes with the recognition site: a and b, without and with the structural waters, respectively; and c, involving the sole TZT anchor. The QC results with the cc-pVTZ(-f) basis set were at this stage limited to the HF level. These ΔE(QC/HF) results are reported without and with the BSSE correction. For illustration, we report in **Figure 19** the compared evolutions of ΔE(QC) and ΔE(SIBFA) in the presence of the structural waters. ΔE(SIBFA) is found to very closely match ΔE(QC), being notably closer to the BSSE-corrected than to the BSSE-uncorrected results. The dizinc core in its complexes with ligands I-V is not a 'rigid' anchorage for the TZT moiety. The Zn-Zn distance varied in the range 4.3 Å (ligand I)-4.85 Å (ligand V).

The QC trends were again closely accounted for and the relative errors were <2 %. Detailed analyses of the separate side-chain interactions with their interacting VIM-2 residues were provided in Ref.[125].

### *Non-enzymatic Zn-proteins. Zn-fingers.*

In an early paper we reported simulations on retroviral 18-residue Zn-fingers with a CCHC core and its CCHH mutant, totaling 173 and 178 atoms, respectively. This was done first by a kinetic Monte Carlo approach and a potential of mean force followed by SIBFA energy-minimization [126]. The resulting structures were compared to the NMR-derived ones. The interaction energies were computed in parallel by SIBFA and QC calculations on six models of the CCHC Zn-finger and four models of the CCHH Zn-finger. For both Zn-fingers, the ranking of $\Delta E_{tot}$(SIBFA) was the same as the $\Delta E_{tot}$(QC) one.

Retroviral Zn-fingers constitute a target for ligands acting as Zn-ejectors by alkylating one of the Zn-coordinating Cys residues. Thus 2-mercaptobenzamide thioesters were designed, synthesized and tested at the National Institute of Health (NIH). In a joint paper with NIH, we performed SIBFA energy minimization on the complex of one such ligand with the nucleocapsid of HIV-1 to optimize the non-covalent interactions prior to acyl transfer [28] . The complex was stabilized by several intermolecular interactions (**Supp. Info 4**): a stacking interaction between the benzyl ring and the indole ring of $Trp_{37}$; a bidentate H-bonding interaction of a terminal formamide and the side-chain of $Gln_{45}$; a H-bond between the carbonyl O of the mercaptocarboxamide and the NH hydrogen of $Gln_{45}$, that is trans to its CO oxygen; and a H-bond between the mercaptosulfur and the main-chain NH hydrogen of $Trp_{37}$. Such interactions enable to anchor the C atom of the mercaptocarboxamide at a proper distance (3.4 Å) to the coordinating S atom of $Cys_{36}$ so that its alkylation could occur in a subsequent step. It would be rewarding to resume studies on Zn-finger-extractor complexes on a more extended scale, with the availability of MD with polarizable potentials, possibly followed by QM/MM modeling of the acyl transfer reaction. A recent MD simulation was reported with accelerated sampling techniques on the Zn-finger of NCp7 indicating the likely coexistence in solution of an extended and of a folded conformation [127].

## Conclusions and Perspectives.

Zn(II) proteins exert prominent function in metabolism, such as in cell growth and differentiation, wound healing, etc. However, when overexpressed, they are the cause of several deadly or invalidating diseases.

It is worth considering the use of polarizable potentials in order to improve the prospects of drug design targeting Zn-metalloproteins. In this review, we have addressed an essential issue: how closely could such potentials reproduce the intermolecular interaction energies from QC calculations in a diversity of representative cases? A reliable match to QC would confer credence to subsequent large-scale MD simulations on the entire protein and a 'box' of waters with periodic boundary conditions resorting to such potentials: such simulations are far from amenable to QC calculations. We focused in this review on the SIBFA potential which was developed and refined in our Laboratories, but similar tests could be carried out as well with the other reviewed potentials. For that purpose, the structures of the tested complexes were made available in the Journals where they were published, and several are available from the Authors if requested.

We started with polyligated Zn(II) complexes found in metallo-enzymes. We could in such complexes unravel the energy contributions at the origin of the preferential coordination number, which could vary between 4 and 6. The second-order contributions, polarization and charge-transfer, were in most cases those leading to the preferred value of n. We also ensured for a proper control of non-additivity of $E_{pol}$ and $E_{ct}$ in polyligated Zn(II) complexes, an issue rarely addressed by others.

Subsequently, essential validations bore on the complexes of the inhibitors with the recognition site of the target proteins. At the outcome of EM on the protein complexes with a series of inhibitors, protein recognition site were isolated with the inhibitor. Single point SIBFA and QC calculations were then performed. Over the years, the size of such complexes varied from 120 [99] to 300 [125], and so did the size of the QC basis sets used for the validation, the largest one being now cc-pVTZ. The latest validations also included up to fourteen structural water molecules, some of which were found to have very large dipole moments. In all simulations, $\Delta E$(SIBFA) and $\Delta E_{tot}$(SIBFA) were shown to reliably match $\Delta E$(QC/HF) and $\Delta E$(QC/DFT-d) or $\Delta E$(QC/MP2).

Another distinctive feature of polarizable potentials related to structural waters. Validations by QC were reported in the case of the Zn(II)/Cu(I) Superoxide Dismutase (SOD) regarding their stabilization energies on an extended model of SOD and also regarding their electrostatic properties, notably the very large dipole moments (permanent plus induced) computed for some [112]. In the case of phosphomannoisomerase (PMI), they were found to favor the preferential binding of a malonate rather than a phosphate mannose surrogate [114] .

The immediate prospects of applications of MD are with the AMOEBA potential, as a follow-up from previous papers [78]. It could be desirable to benchmark this potential as well against QC. A highly desirable step forward regarding the SIBFA potential integrated in the Tinker-HP code is the portage of its CPU code to GPUs. This could indeed very significantly boost this potential for much larger-scale applications than reported so far. Continuum solvation could also be used.[128]

Concurrently, benchmarking the accuracy of machine learning foundation models for atomistic molecular dynamics against high-level QC calculations offers an exciting avenue for exploration [129], especially as their speed improves. [130-131]

Could the integration of these data-driven potentials define the next frontier in Zn-metalloprotein modeling? Could we directly express the SIBFA contribution from these models? Incoming works will answer these questions.

## Acknowledgments

We are grateful to our colleagues in both theoretical and experimental chemistry who made the results reported in this review possible and co-authored the cited papers. The theoretical research since 1995 was conducted in close collaboration with co-workers across several international institutions, including those in France (Paris, École Normale Supérieure de Cachan, and Montpellier), the United States (Rockville, New York, and Chapel Hill), and Denmark (Copenhagen). We also wish to acknowledge our long-standing, fruitful collaboration with experimentalists at the Laboratoire de Chimie Bio-Organique et Bio-Inorganique (Université Paris-Saclay) and the Institut des Biomolécules Max Mousseron at the University of Montpellier. Finally, the theoretical computations were made possible through generous allocations of supercomputing time on the GENCI high-performance computing facilities of IDRIS in Orsay, CINES in Montpellier, and CRIANN mesocentre in Rouen.

**Captions for figures.**

**Figure 1**. Molecular structures of L and D isomers of the thiomandelate and captopril Zn-metalloenzyme inhibitors. Reprinted from Ref. 99 with permission.

**Figure 2**. Thiomandelate and captopril complexes with a beta-lactamase binding site. $\Delta E_{tot}$(SIBFA) vs. ΔE(MP2) and ΔE(DFT) intermolecular interaction energies. Reprinted from Ref. 99 with permission.

**Figure 3.** X-ray structure of L1 bacterial Zn-metalloenzyme complexed by a triazole-thione inhibitor. In the $(His)_3$ site (left), the three His residues from top-to-bottom are labelled H116, H118, and H196 respectively. In the ((Asp-$(His)_2$) site (right), the residues are labelled D120, H121, and H263 respectively.

.

**Figure 4.** Representation of the complex of the triazole thione Zn(II)-binding group in the dizinc binding site of the bacterial L1 Zn-metallo-beta-lactamase. Reprinted from Ref. 106 with permission. In the $(His)_3$ site (left), the three His residues from top-to-bottom are labelled H116, H118, and H196 respectively. In the ((Asp-$(His)_2$) site (right), the residues are labelled D120, H263 and H121, respectively.

**Figure 5.** Compared evolutions of ΔE(QC/RVS) and ΔE(SIBFA) and their separate contributions in the TZT-ligated and unligated mono- and dizinc L1 binding site. Reprinted from Ref. 106 with permission

**Figure 6.** Representation of the optimized complexes of three TZT derivatives with the dizinc binding site of VIM-2 Zn-metallo-beta-lactamase. Reprinted from Ref.106 with permission,
In the $(His)_3$ site (left), the three His residues from top-to-bottom are labeled H116, H118, and H196 respectively. In the (Asp-(His-Cys site) (right), the residues are labeled D120, H263, and C221 respectively.

**Figure 7.** X-ray structure of the Zn(II)/Cu(I) bimetallic binding site of Superoxide Dismutase. Reprinted from Ref. 112 with permission.

**Figure 8.** Representation of the network of discrete water molecules in the bimetallic SOD Zn(II)/Cu(I) binding site bridging Glu131 and Arg141. Water molecules with the highest dipole moments are highlighted. Reprinted from Ref. 112 with permission.

**Figure 9.** Values of the summed dipole moments of the 28-water clusters in complexes a-f. . Reprinted from Ref. 112 with permission.

**Figure 10.** Compared evolutions of ΔE(QC/RVS) and ΔE(SIBFA) and its $E_1$ and $E_2$ contributions in complexes a-f. Reprinted from Ref. 112 with permission.

**Figure 11.** Compared evolutions of ΔE(B97D) and $\Delta E_{tot}$(SIBFA) and of $\Delta E_{corr}$ and $E_{disp}$(SIBFA) in complexes a-f. Reprinted from Ref. 112 with permission.

**Figure 12a.** Compared evolutions of ΔE(QC/RVS) and ΔE(SIBFA) stabilization energies due to the 28-water network and of the corresponding $E_1$ and $E_2$(SIBFA) in complexes a-f.. Reprinted from Ref. 112 with permission.

**Figure 12b.** Compared evolutions of ΔE(DFT/B97D) and $\Delta E_{tot}$(SIBFA) stabilization energies due to the 28-water network. Reprinted from Ref. 112 with permission.

**Figure 13.** Three-dimensional structure of the Zn-metalloenzyme phosphomannose isomerase PMI. .Reprinted from Ref. 114 with permission.

**Figure 14.** Representation of β-M6P, α-M6P, β-G6P, and β-6DCM PMI active site ligands, denoted as 1-4. Reprinted from Ref. 114 with permission.

F

**Figure 15a.** Representation of the network of structured waters in the PMI-4 complex. Reprinted from Ref. 114 with permission

**Figure 15b.** Representation of the network of structured waters in the PMI-1 complex. Reprinted from Ref. 114 with permission.

**Figure 16.** Molecular structures of 5PAH and its three dehydroxylated analogs. Reprinted from Ref. 90 with permission.

**Figure 17a.** Compared evolutions of ΔE(SIBFA) and ΔE(HF) and of $\Delta E_{tot}$(SIBFA) and ΔE(DFT-d) with the CEP 4-31G(2d) basis set. Reprinted from Ref. 90 with permission

**Figure 17b.** Compared evolutions of ΔE(SIBFA) and ΔE(HF) and of $\Delta E_{tot}$(SIBFA) and ΔE(DFT-d) with the ccpvtz basis set. Reprinted from Ref. 90 with permission

**Figure 18.** Molecular structures of the five VIM-2 inhibitors. Reprinted from Ref. 125 with permission.

**Figure 19.** Compared evolutions of ΔE(QC/HF) and ΔE(SIBFA) in an extended recognition site of VIM-2: -the inhibitors bound in the absence (a) and presence (b) of structural waters; (c) the sole TZT extracted from the inhibitors in the five structures. Reprinted from Ref. 125 with permission.

## Supplementary Figures S1-S4.

**Supp. Info 1**. Bidentate vs. mono-dentate complexes of the hydroxamate Zn(II) binding group extracted from the complex of a phosphoramidate inhibitor with the PMI metalloenzyme.

From Figure 10 of El Khoury et al. *J. Comput. Chem***. 2017,** *38*, 1897

**Supp. Info 2**. Representation of three most representative complexes of 5PAH in the extended recognition site of PMI.

From Figure 2(a-c) of Gresh et al., *J. Comput. Chem.* **2016***, 37*, 2770.

**Supp. Info 3.** Representation of the energy-minimized complexes of the five TZT-based inhibitors with the extended dizinc recognition site of VIM2.

From Figure 2 of Kwapien et al., *J. Comput. Chem* **2021***, 42*, 86.

**Supp. Info 4.** Complex between the NCp7 Zn-finger and a mercaptocarboxamide ligand. From Jenkins et al., *J. Am. Chem. Soc.* **2007,** *129*, 11067.

**Figure 1**. Molecular structures of L and D isomers of the thiomandelate and captopril Zn-metalloenzyme inhibitors.

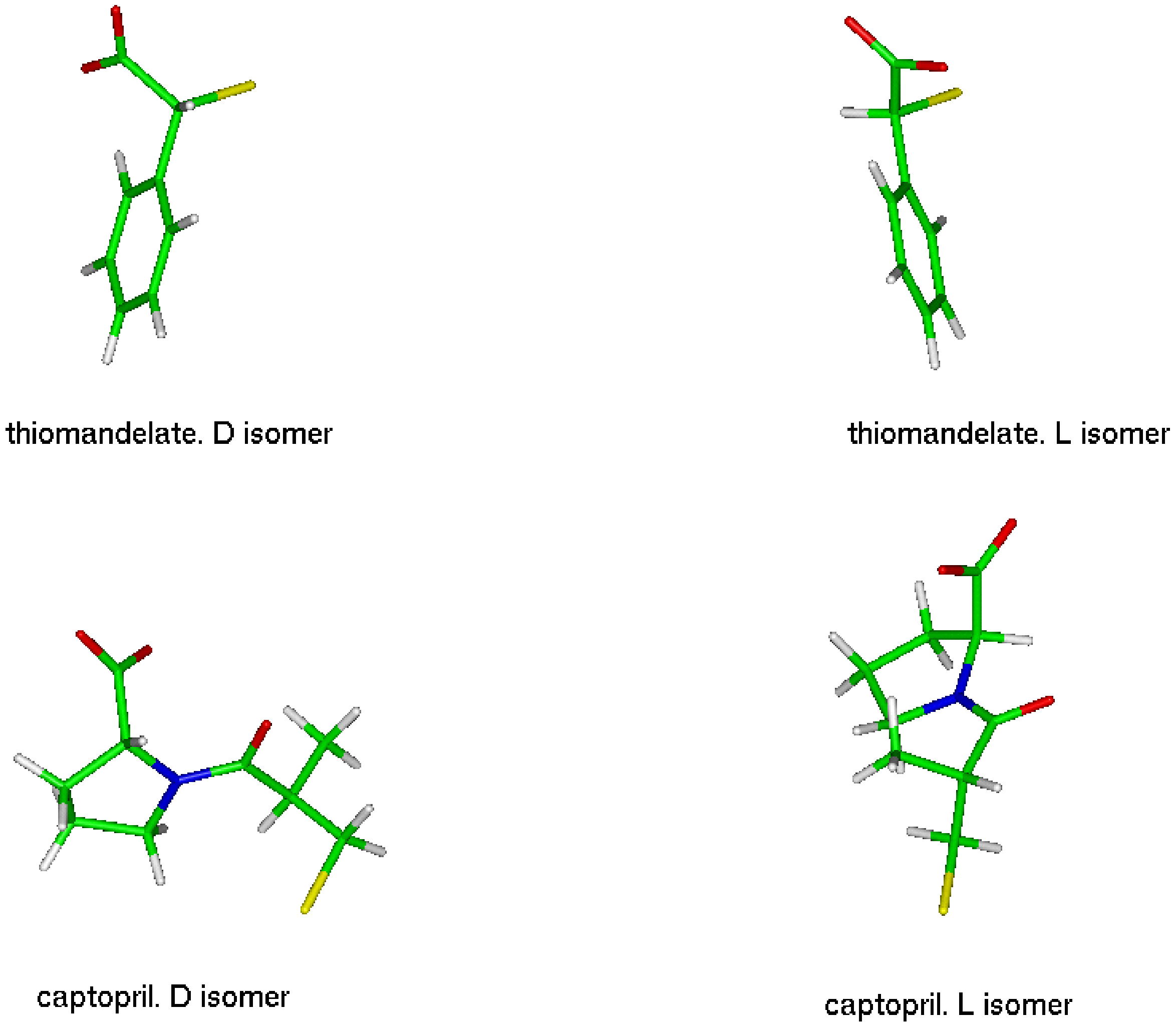

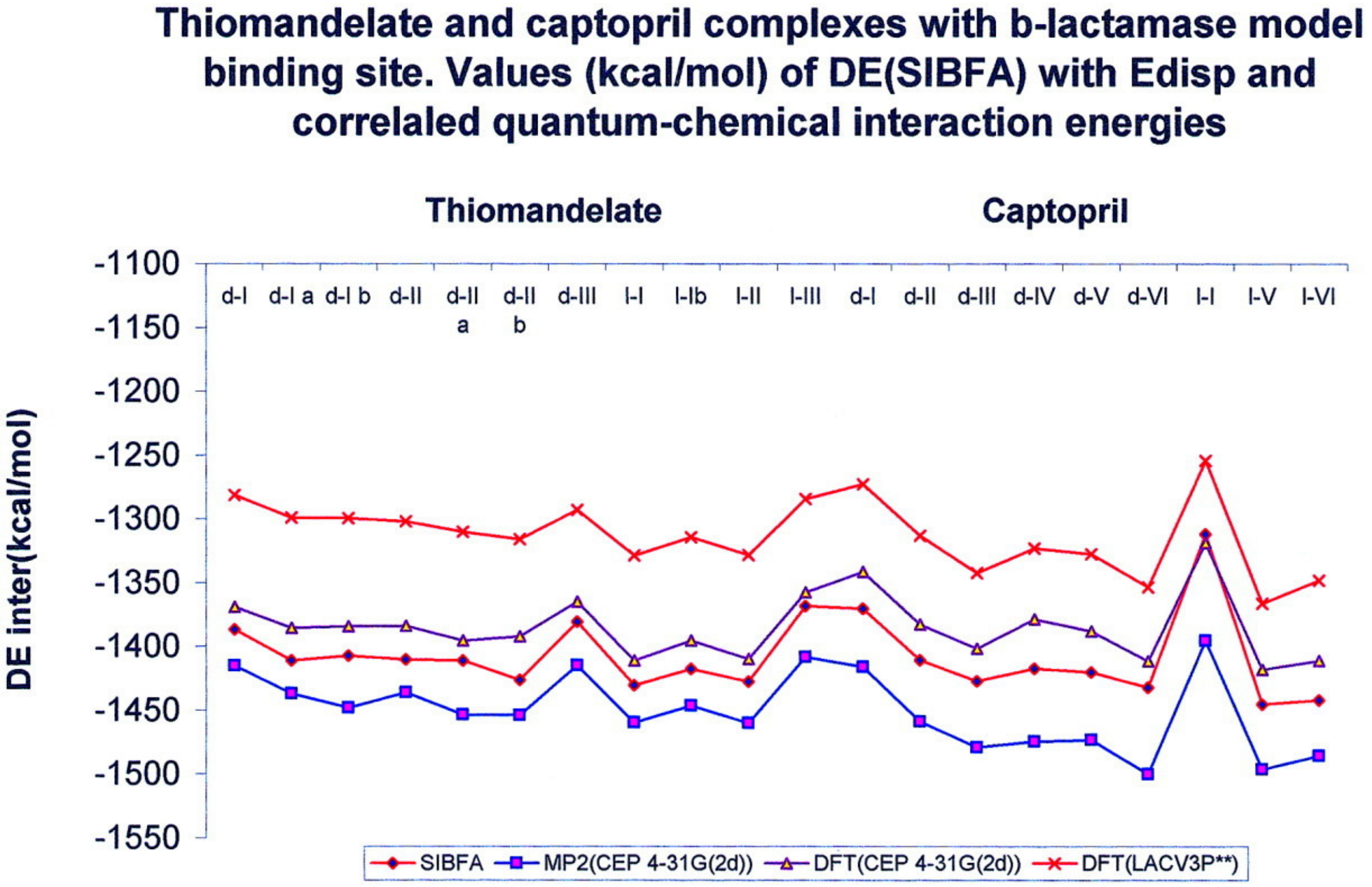

**Figure 2**. Thiomandelate and captopril complexes with a beta-lactamase binding site. ΔEtot(SIBFA) vs. ΔE(MP2) and DE(DFT) intermolecular interaction energies

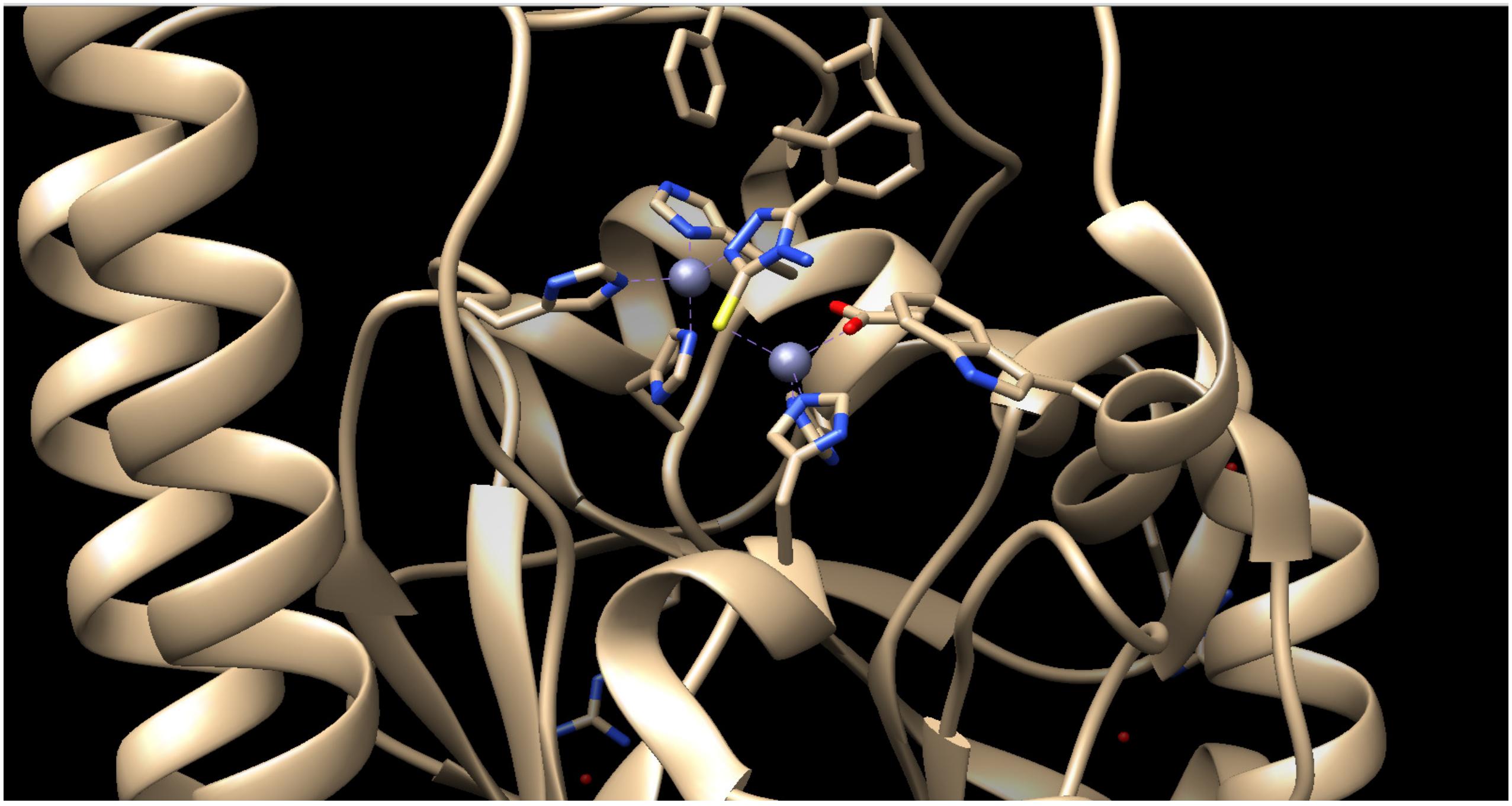

**Figure 3.** X-ray structure of L1 bacterial Zn-metalloenzyme complexed by a triazole-thione inhibitor. In the $(His)_3$ site (left), the three His residues from top-to-bottom are labeled H116, H118, and H196 respectively. In the $((Asp-(His)_2)$ site (right), the residues are labeled D120, H121, and H263 respectively.

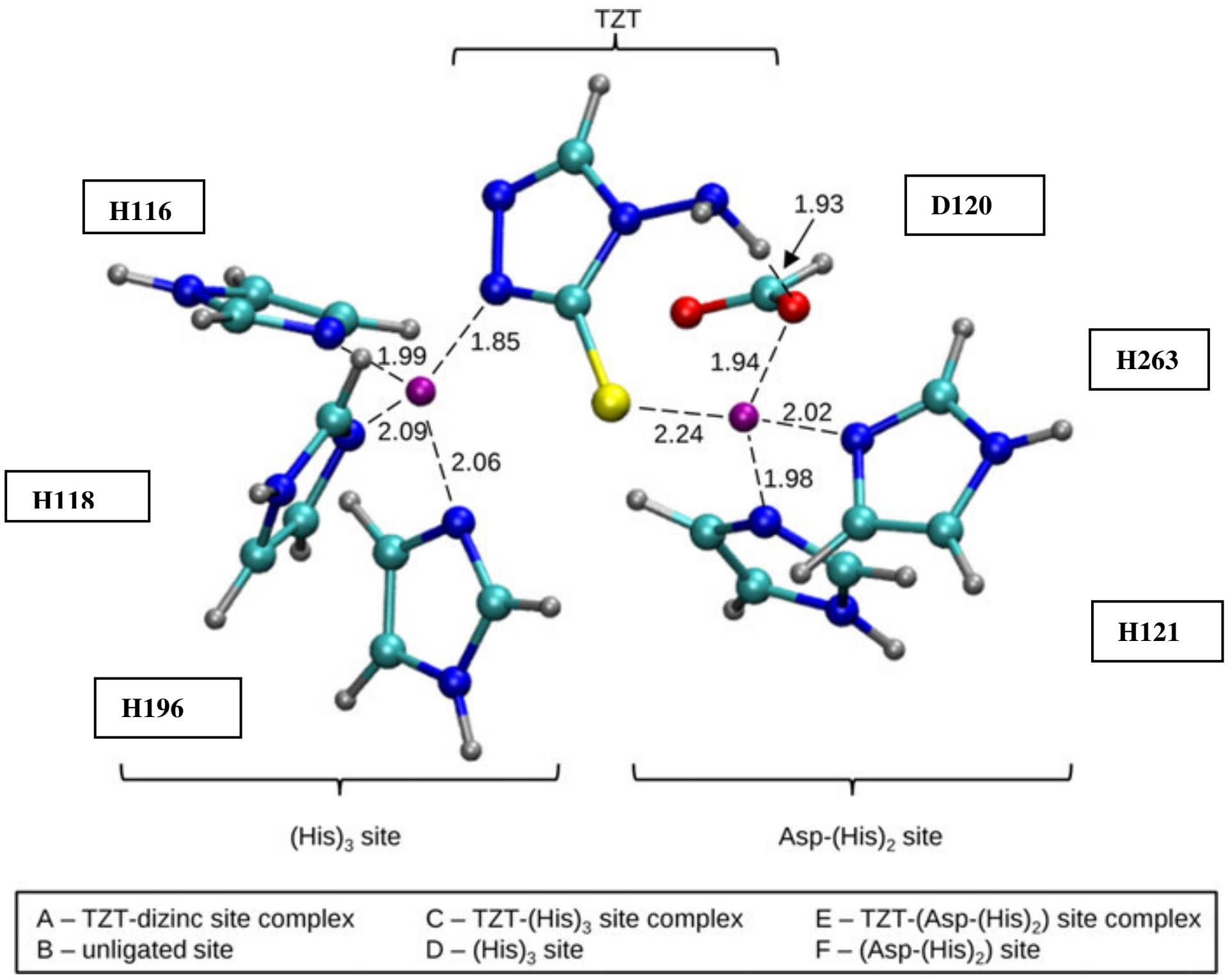

**Figure 4.** Representation of the complex of the triazole thione Zn(II)-binding group in the dizinc binding site of the bacterial L1 Zn-metallo-beta-lactamase.

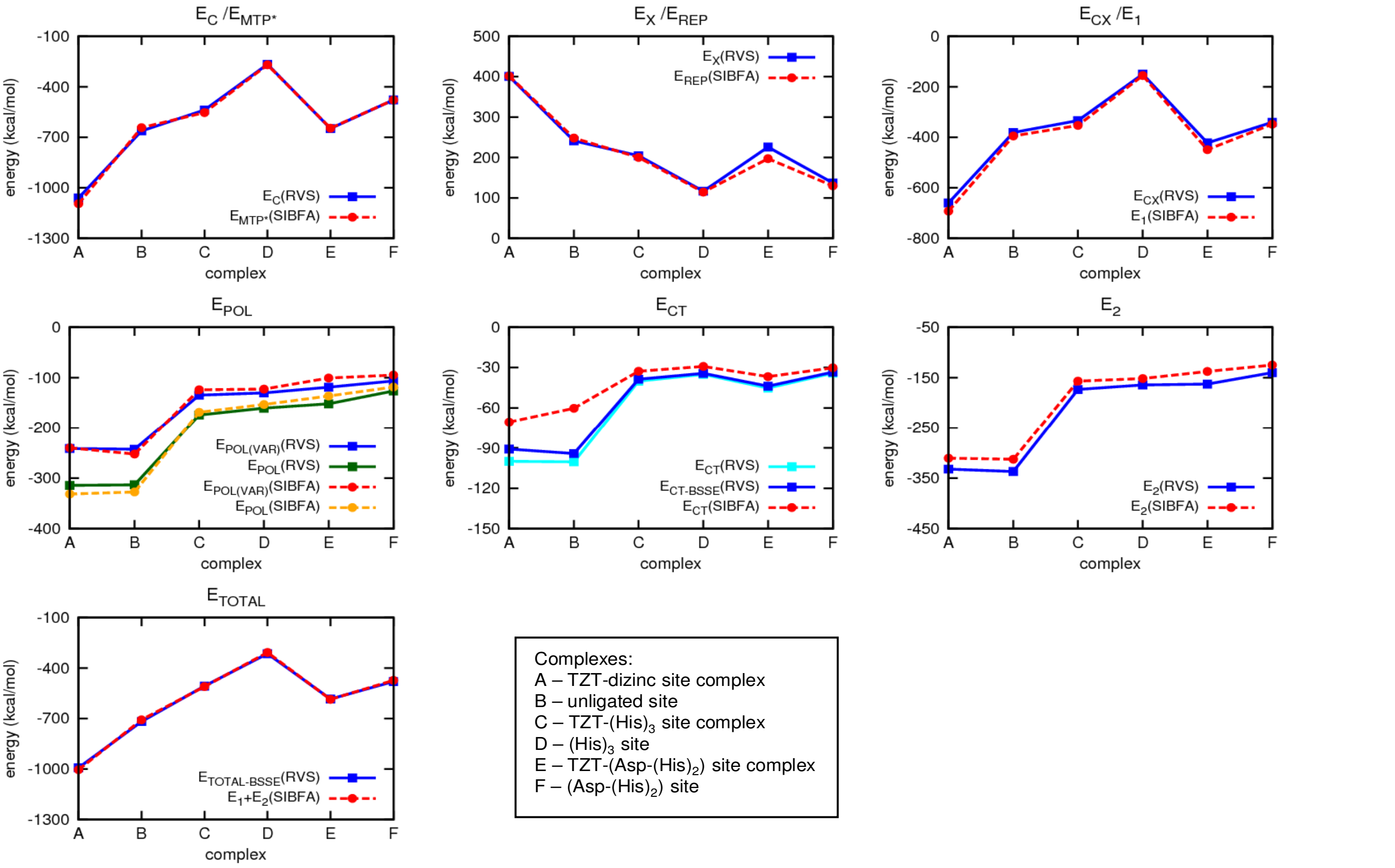

**Figure 5.** Compared evolutions of ΔE(QC/RVS) and ΔE(SIBFA) and their separate contributions in the TZT-ligated and unligated mono- and dizinc L1 binding site.

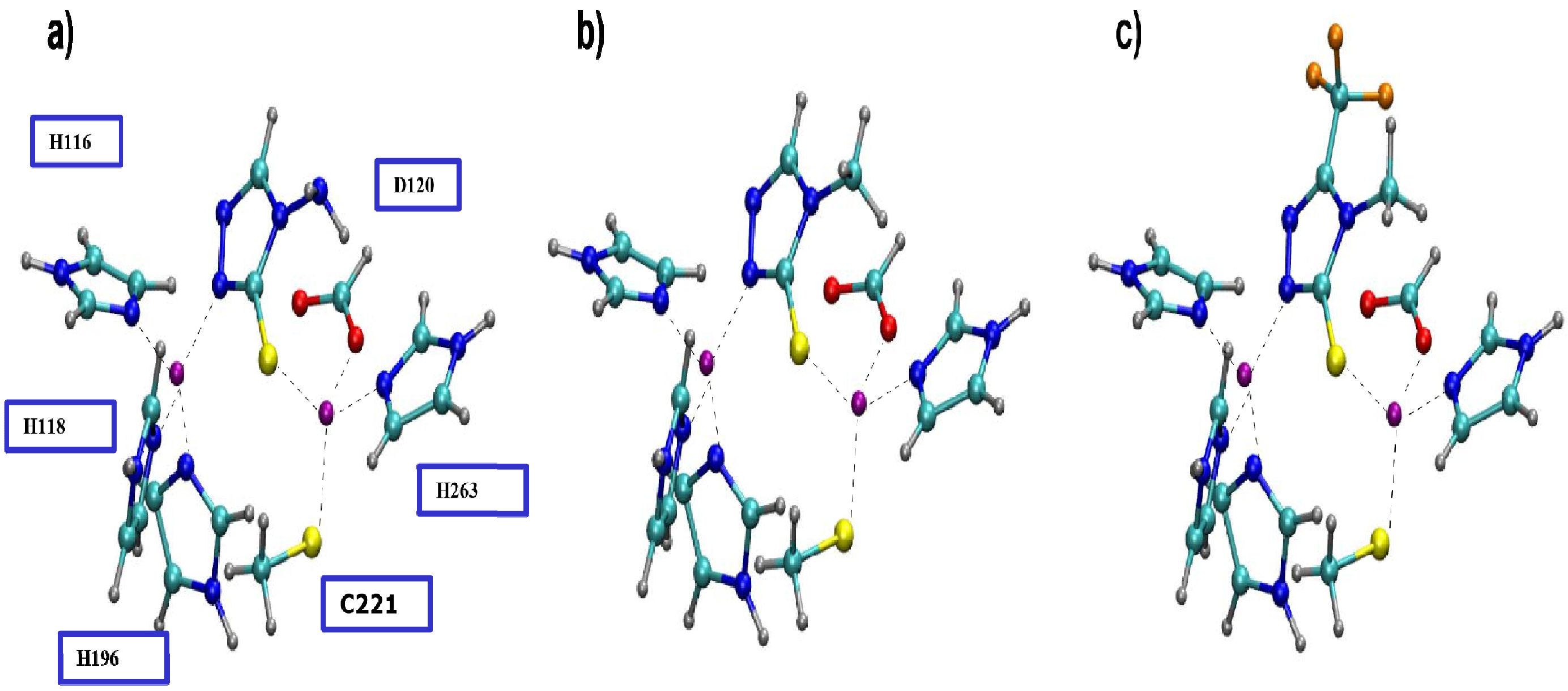

**Figure 6.** Representation of the optimized complexes of three TZT derivatives with the dizinc binding site of VIM-2 Zn-metallo-beta-lactamase. In the $(His)_3$ site, the three His residues from top-to-bottom are labeled H116, H118, and H196 respectively. In the (Asp-(His-Cys site), the residues are labeled D120, H263, and C221 respectively.

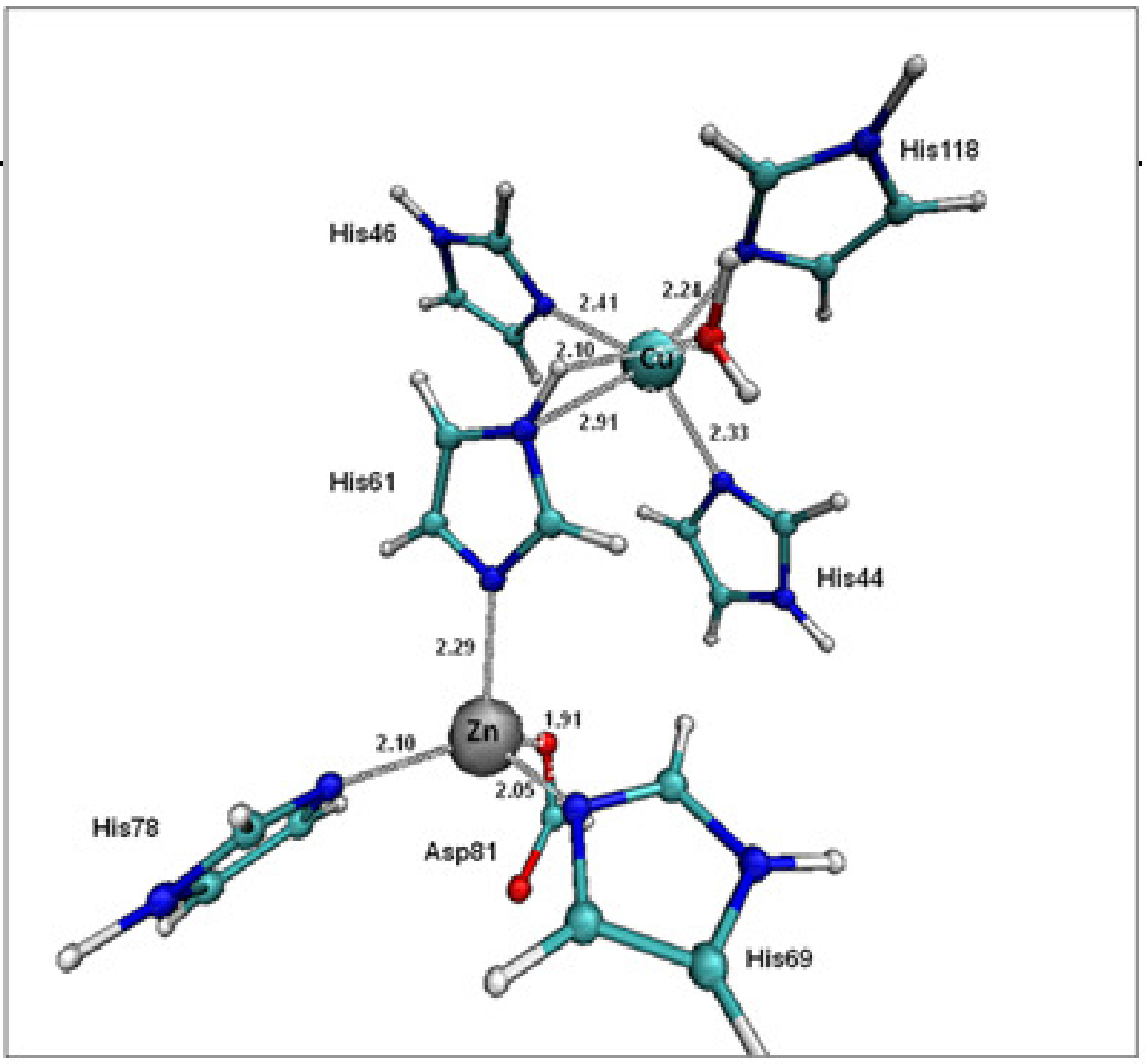

**Figure 7.** X-ray structure of the Zn(II)/Cu(I) bimetallic binding site of Superoxide Dismutase.

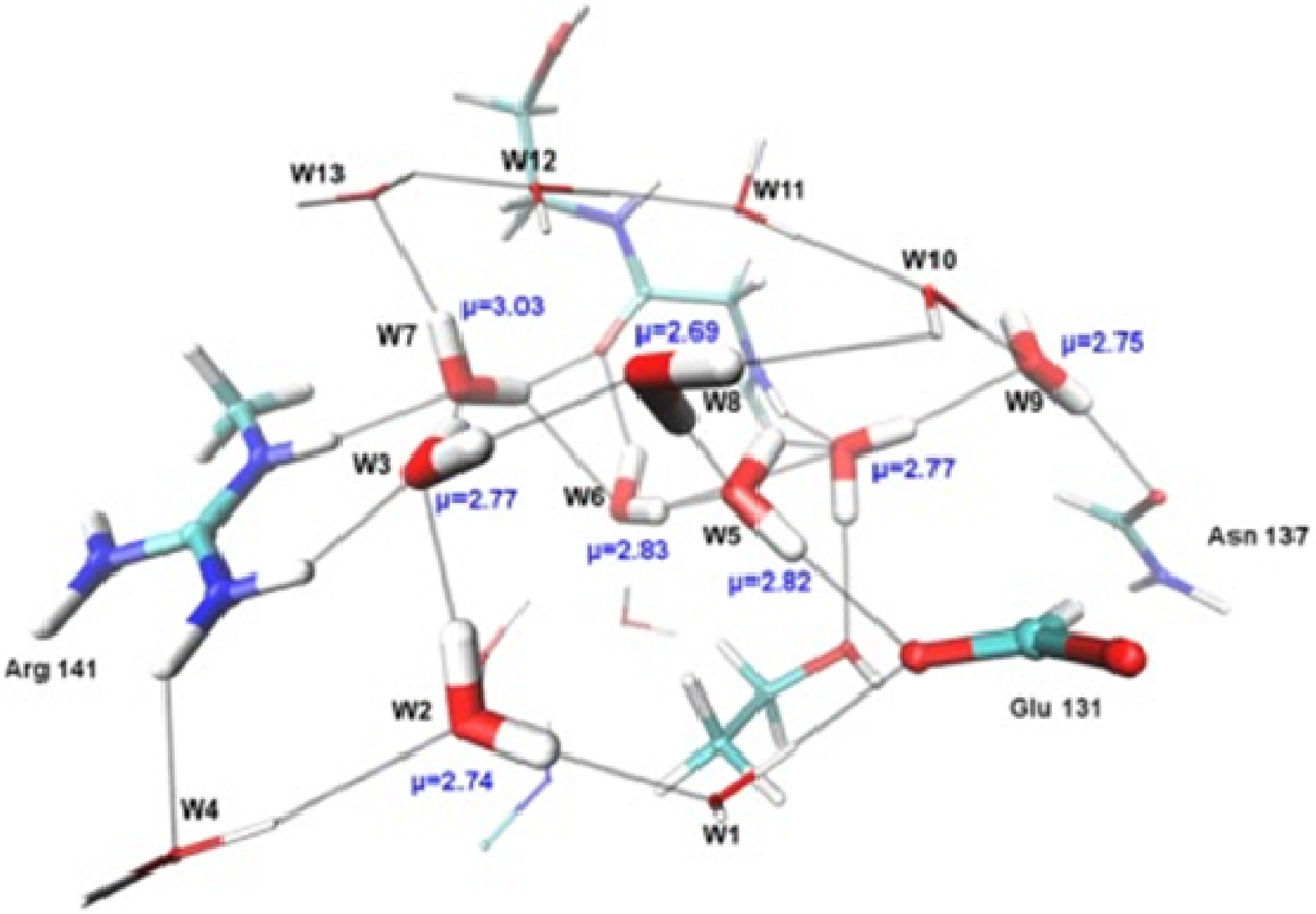

**Figure 8.** Representation of the network of discrete water molecules in the bimetallic SOD Zn(II)/Cu(I)binding site bridging Glu131 and Arg141.
Water molecules with the highest dipole moments are highlighted.

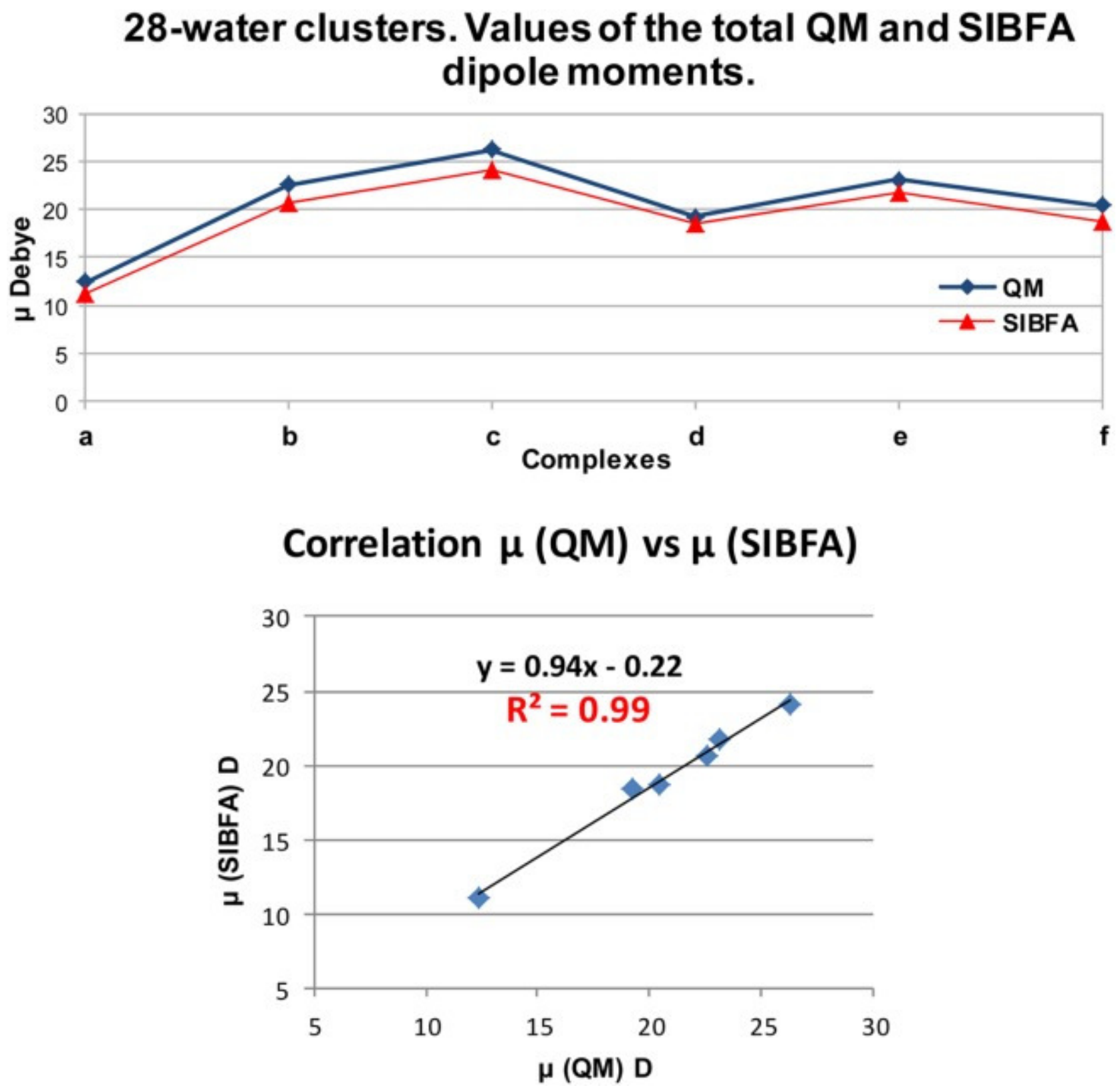

**Figure 9.** Values of the summed dipole moments of the 28-water clusters in complexes a-f.

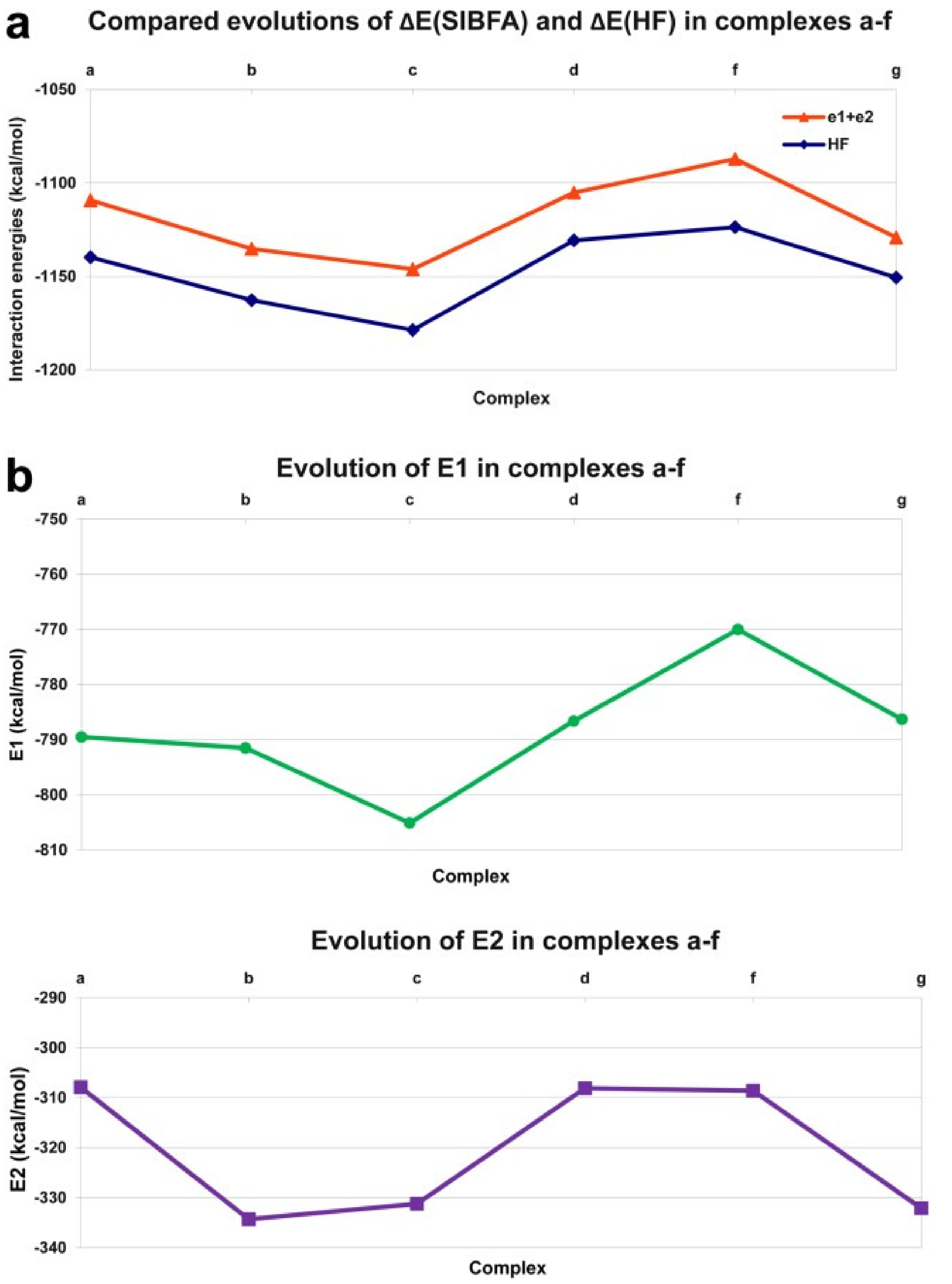

**Figure 10.** Compared evolutions of ΔE(QC/RVS) and ΔE(SIBFA) and its $E_1$ and $E_2$ contributions in complexes a-f

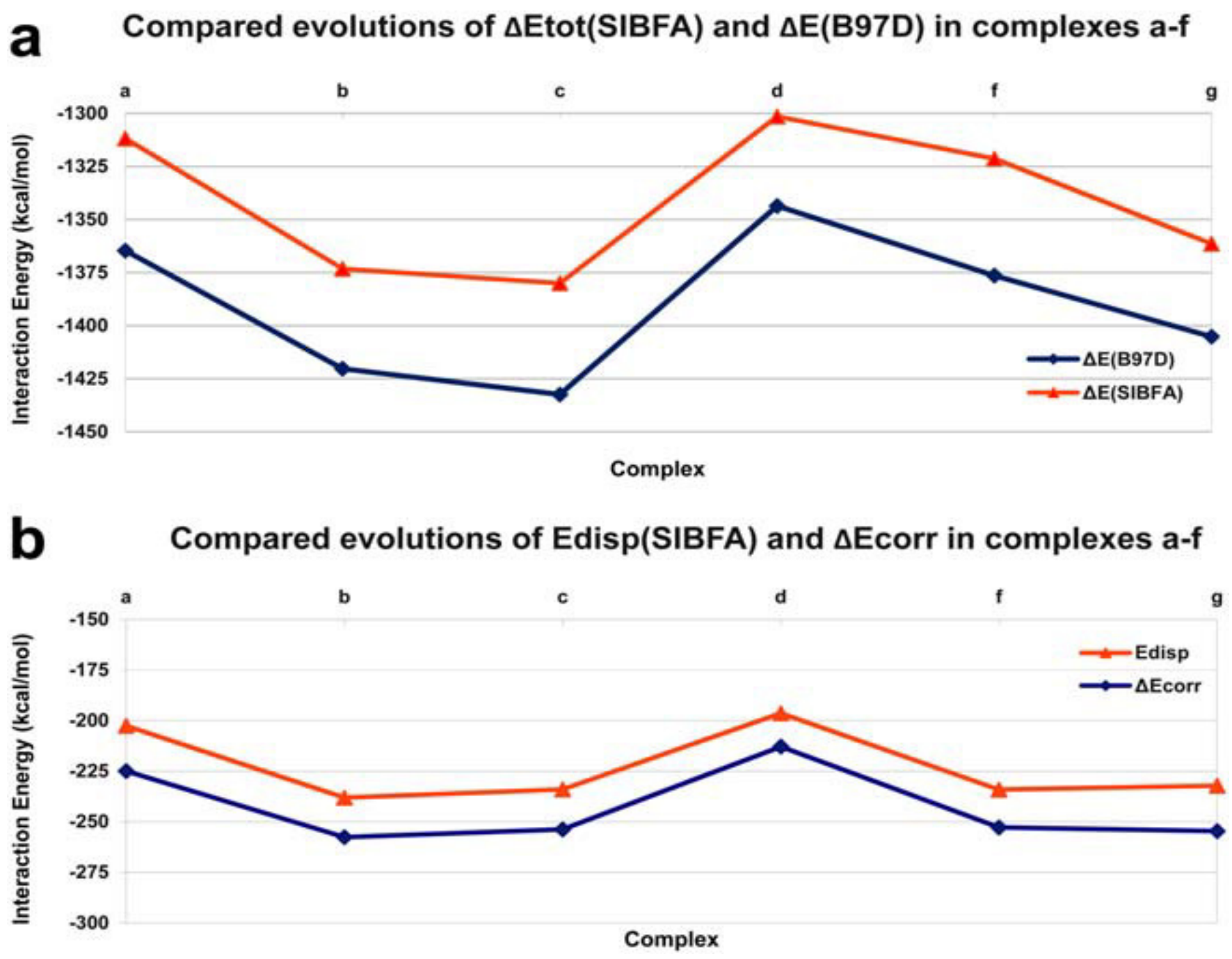

**Figure 11.** Compared evolutions of ΔE(B97D) and $\Delta E_{tot}$(SIBFA) and of $E_{corr}$ and $E_{disp}$(SIBFA) in complexes **a-f**.

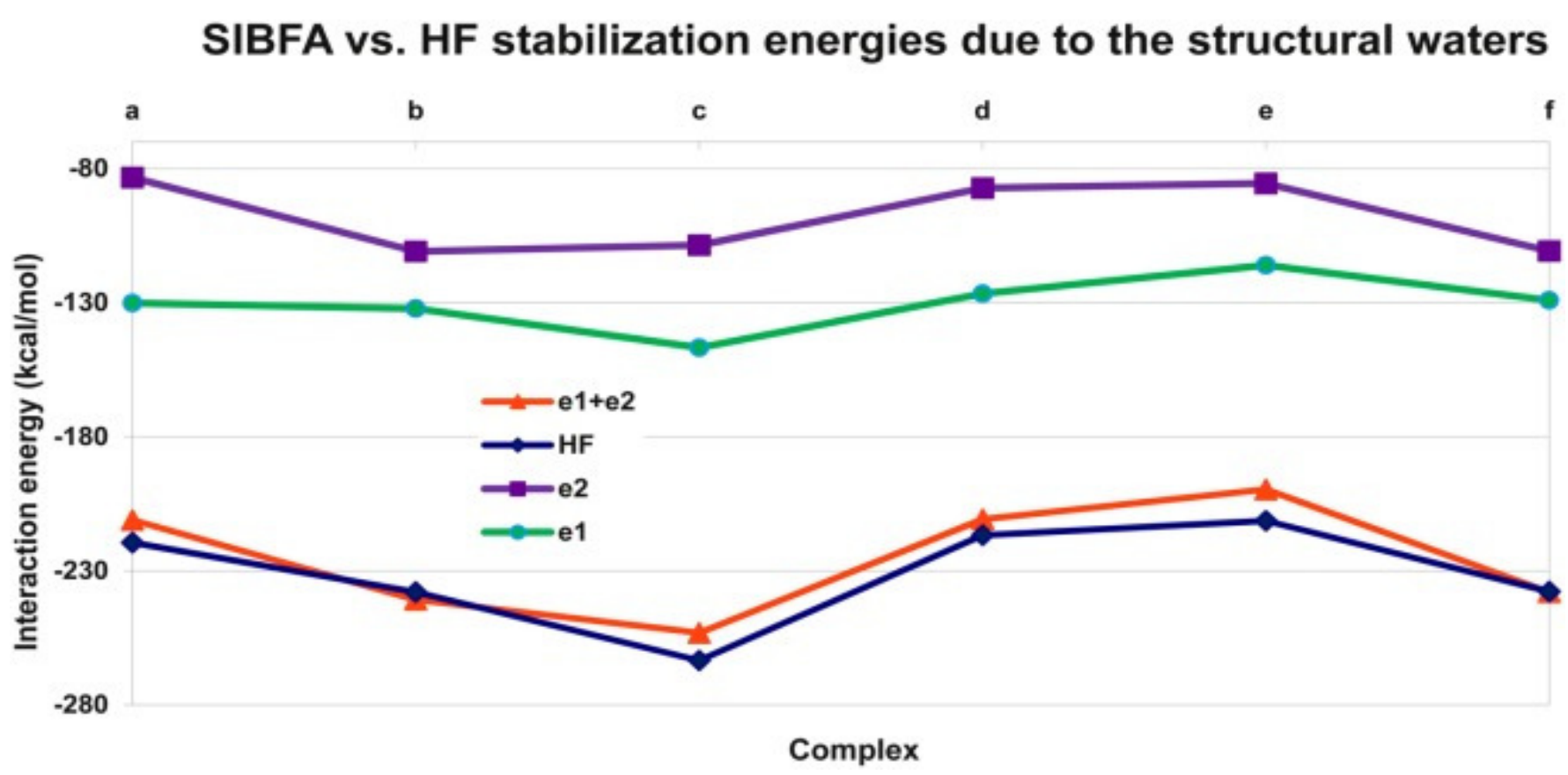

**Figure 12a.** Compared evolutions of ΔE(QC/RVS) and ΔE(SIBFA) stabilization energies due to the 28-water network and of the corresponding $E_1$ and $E_2$(SIBFA) in complexes a-f.

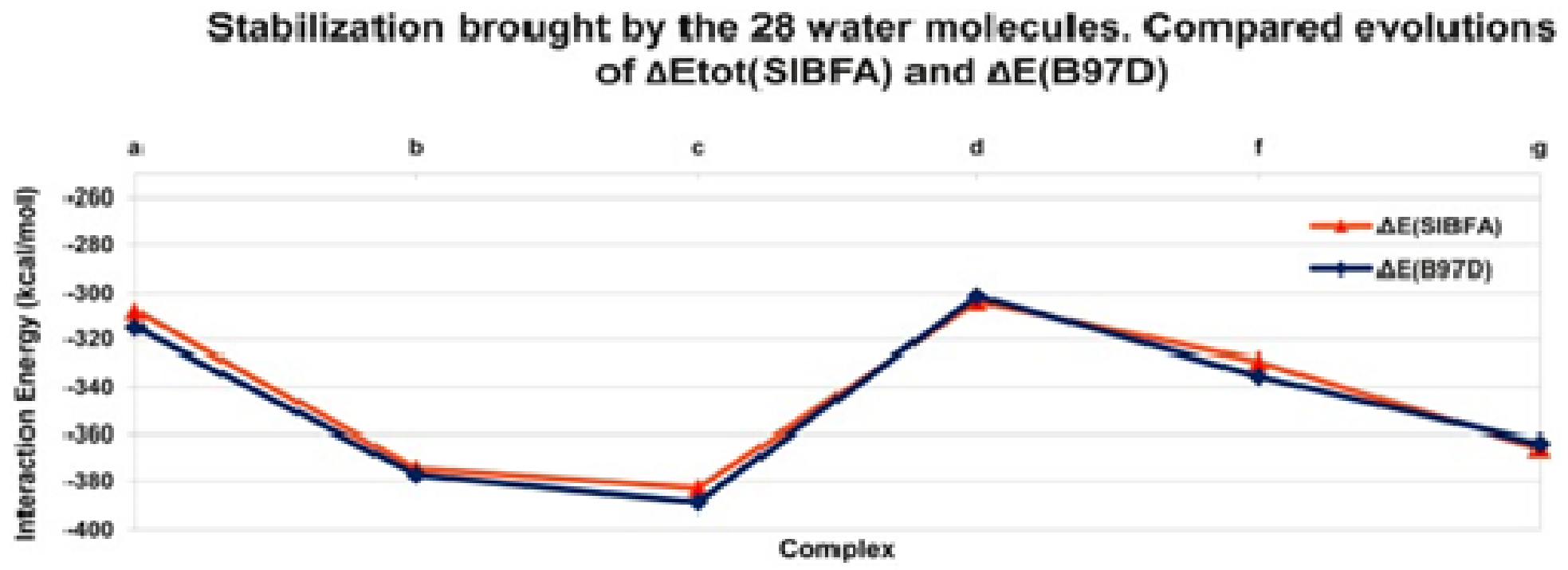

**Figure 12b.** Compared evolutions of ΔE(DFT/B97D) and $\Delta E_{tot}$(SIBFA) stabilization energies due to the 28-water network.

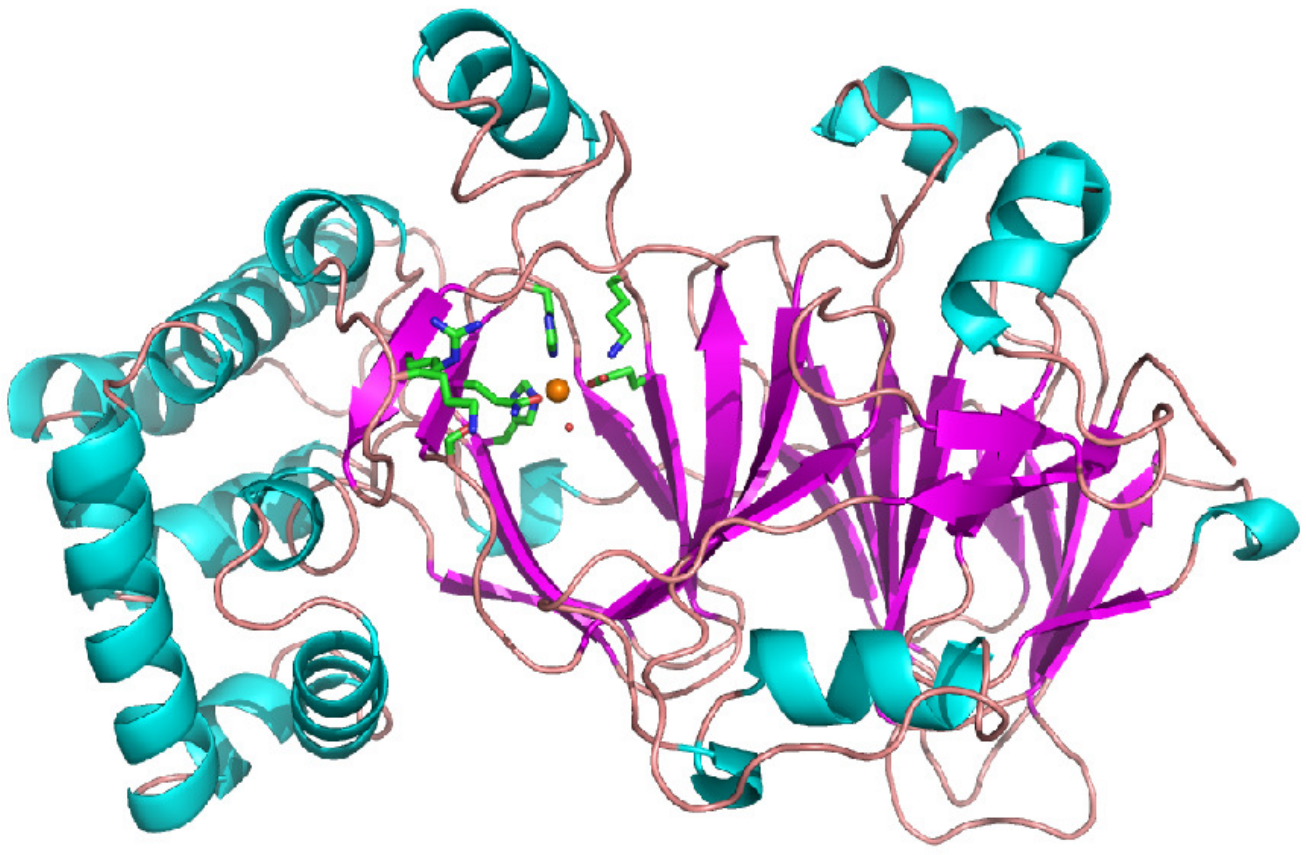

**Figure 13.** Three-dimensional structure of Zn-metalloenzyme phosphomannose isomerase PMI.

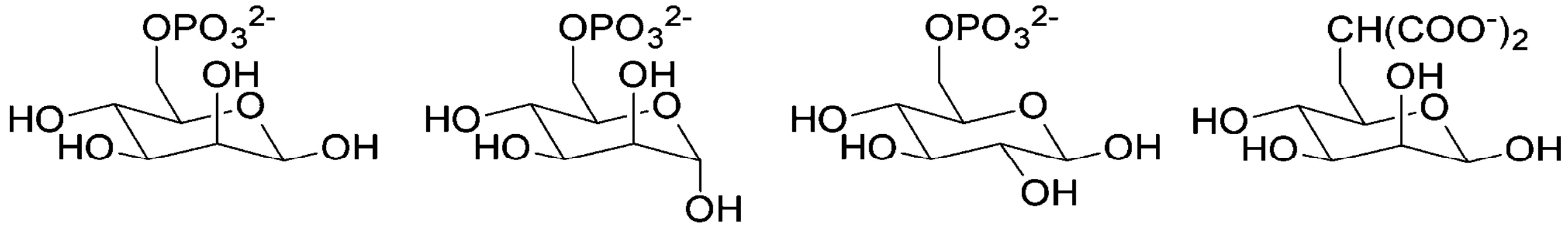

β-M6P **1** α-M6P **2** β-G6P **3** β-6DCM **4**

**Figure 14.** Representation of β-M6P, α-M6P, β-G6P, and β-6DCM PMI active site ligands, denoted as **1-4**.

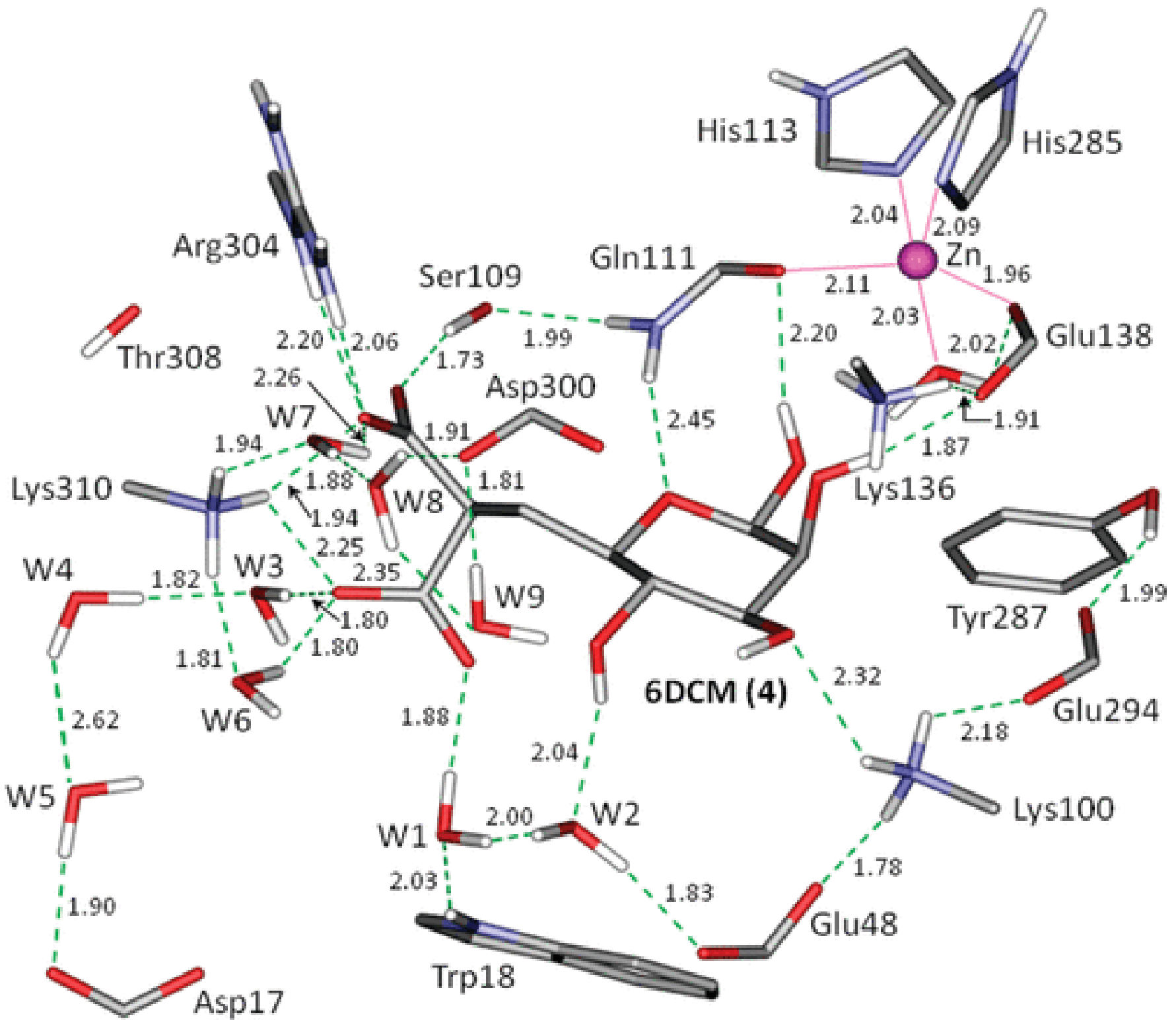

**Figure 15a.** Representation of the network of structured waters in the PMI-**4** complex

**Figure 15b.** Representation of the network of structured waters in the PMI-**1** complex.

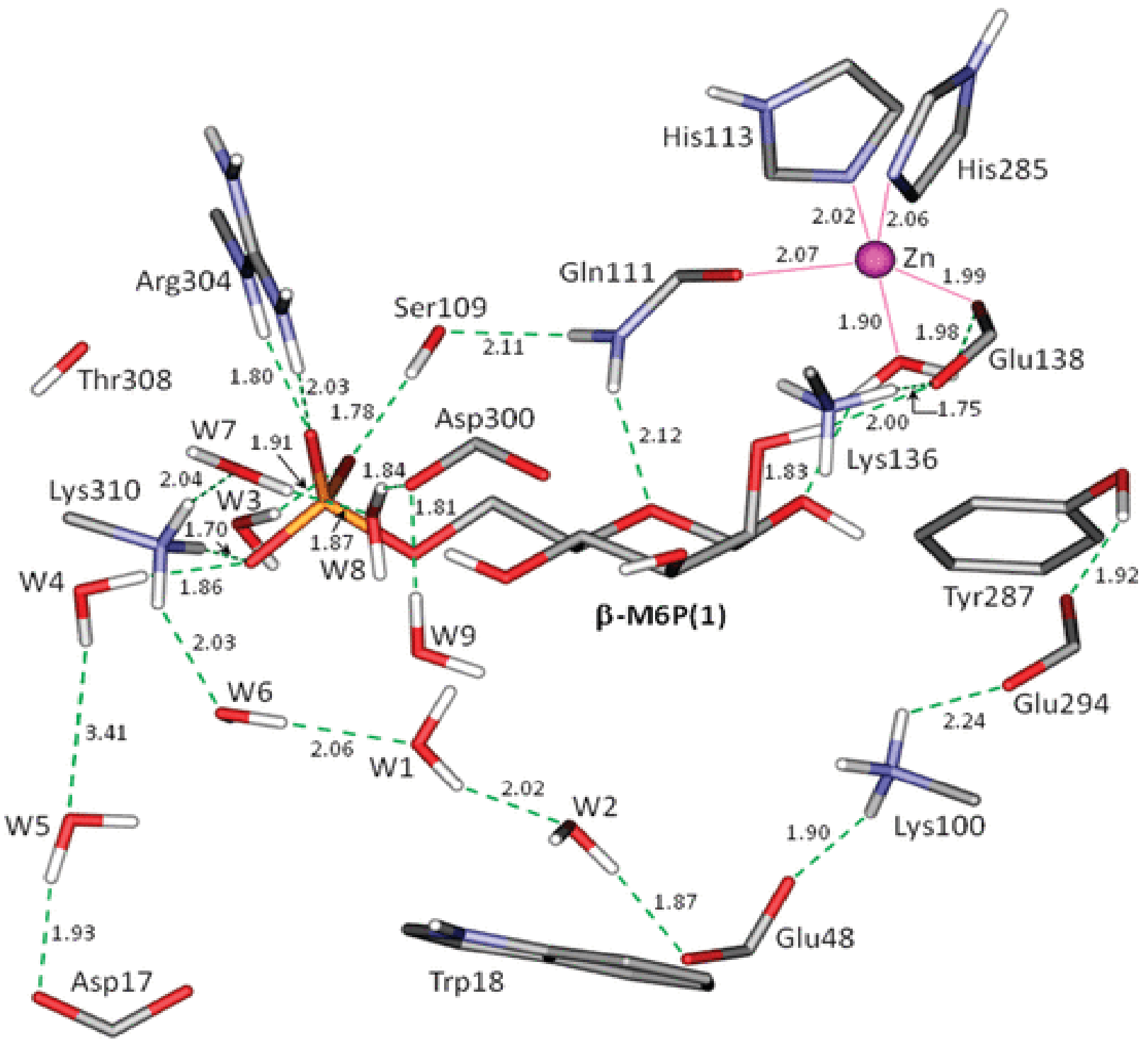

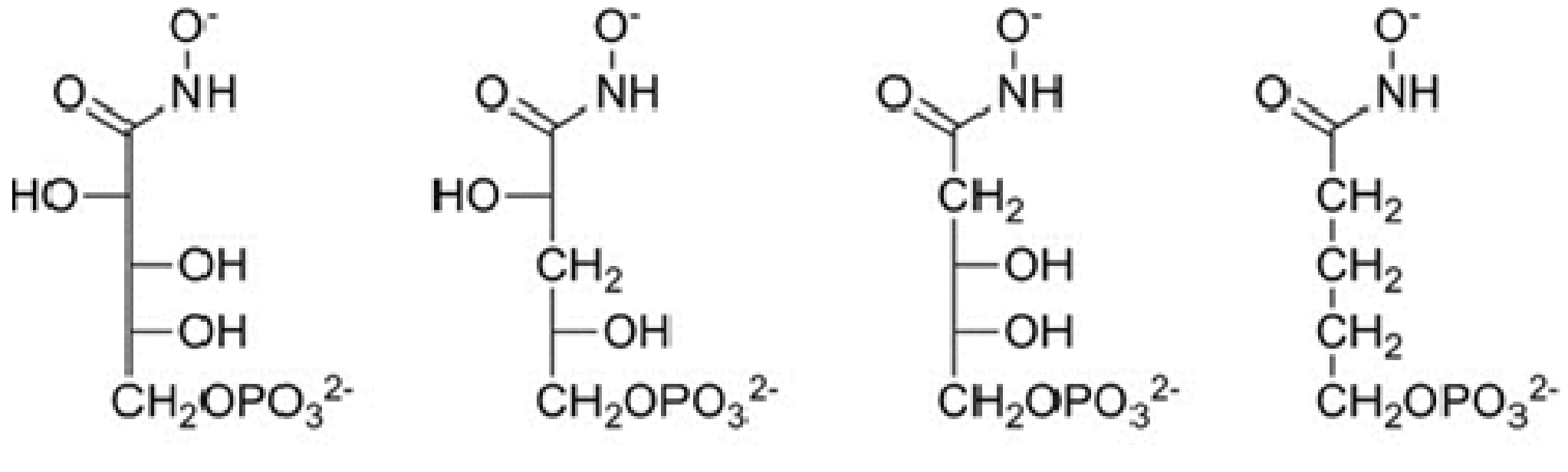

5PAH **a** | 3d-5PAH **b** | 2d-5PAH **c** | 5PPH **d**

**Figure 16.** Molecular structures of 5PAH and its three dehydroxylated analogs.

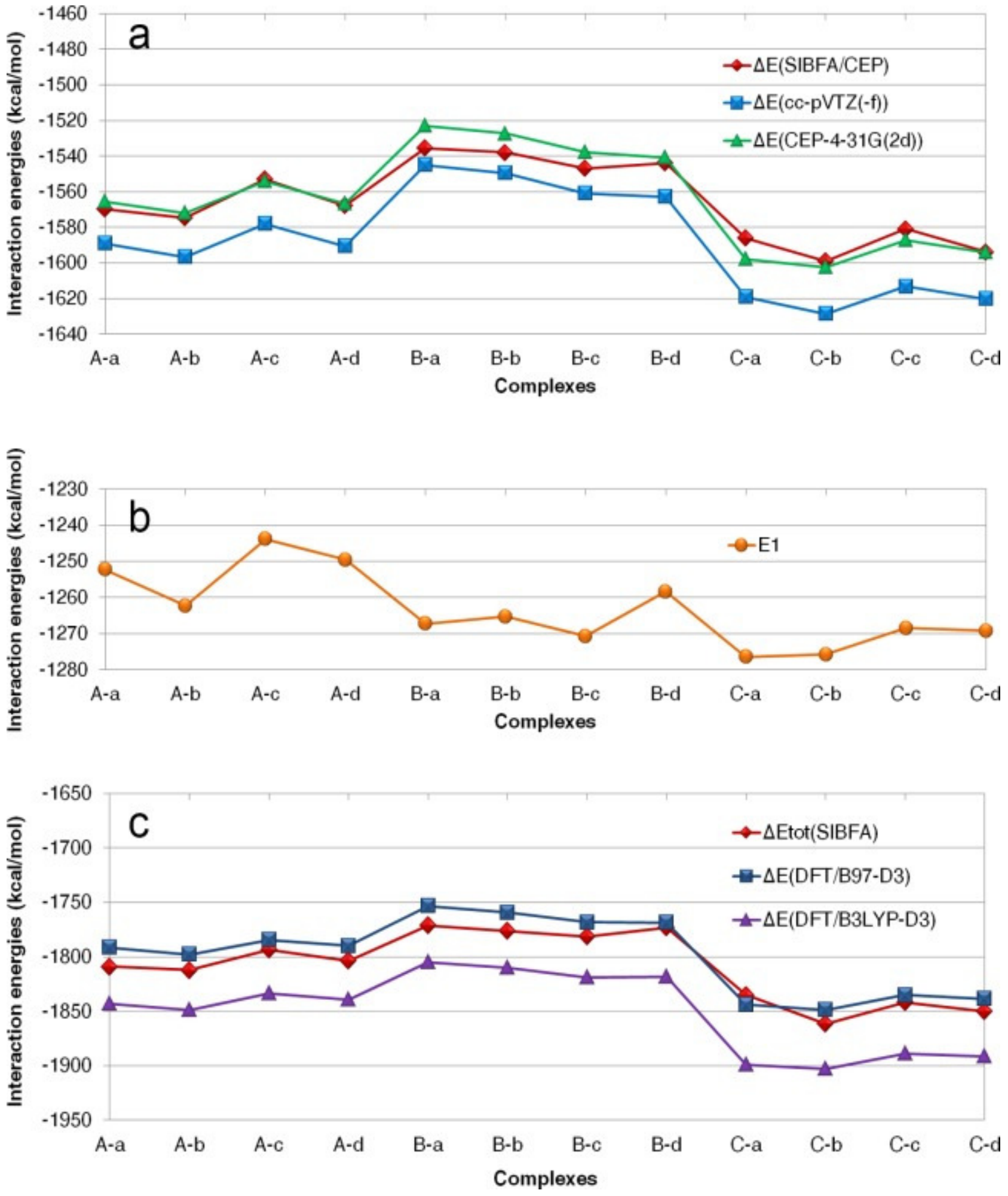

**Figure 17a.** Compared evolutions of ΔE(SIBFA) and ΔE(HF) and of $\Delta E_{tot}$(SIBFA) and ΔE(DFT-d) with the CEP 4-31G(2d) basis set.

**Figure 17b.** Compared evolutions of ΔE(SIBFA) and ΔE(HF) and of $\Delta E_{tot}$(SIBFA) and ΔE(DFT-d) with the cc-pvtz basis set

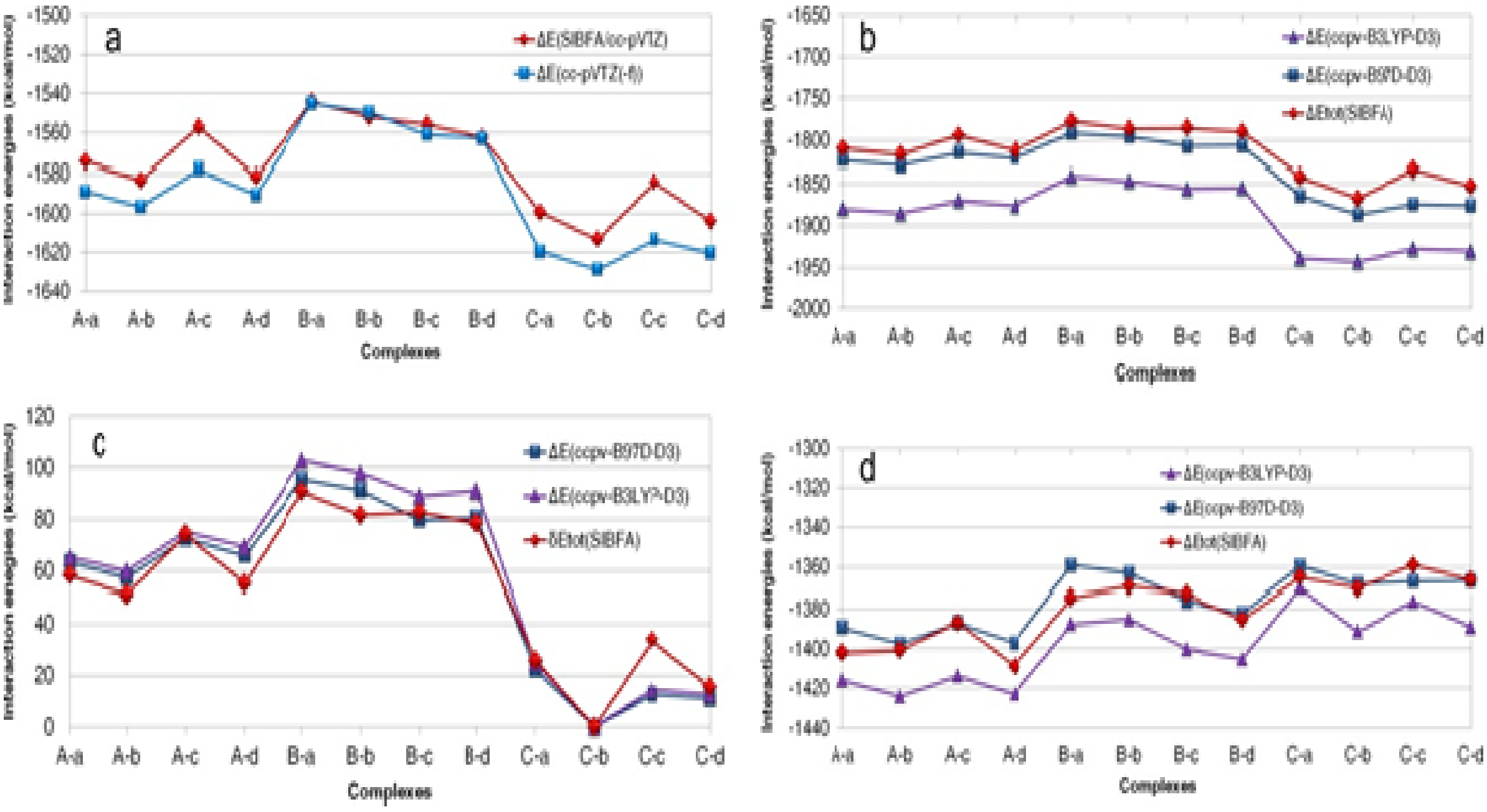

**Figure 18.** Molecular structures of the five VIM2 inhibitors.

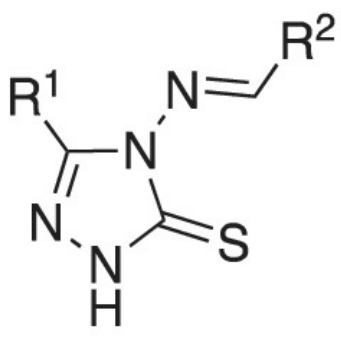

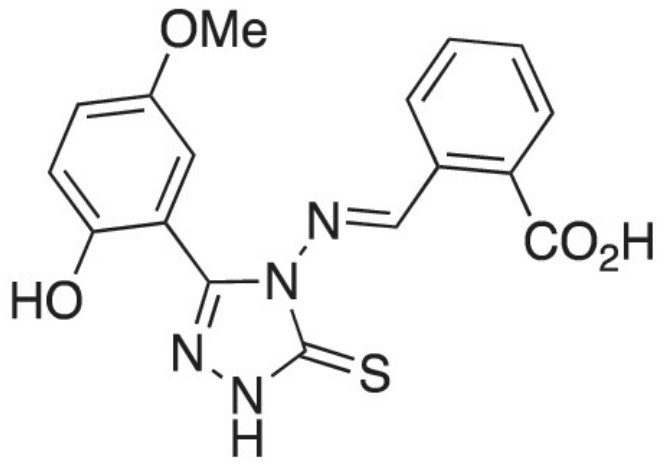

JMV4690 (I)

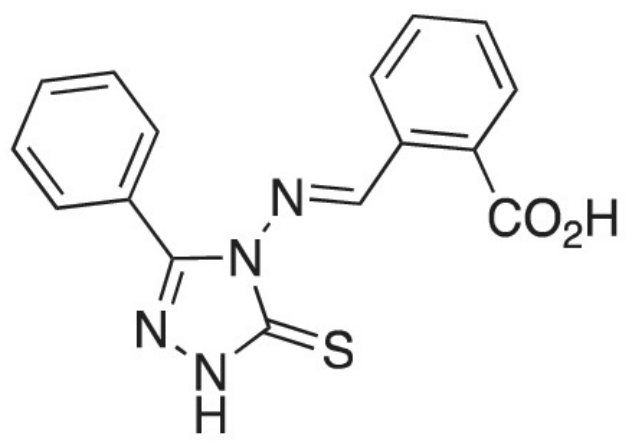

JMV4390 (II)

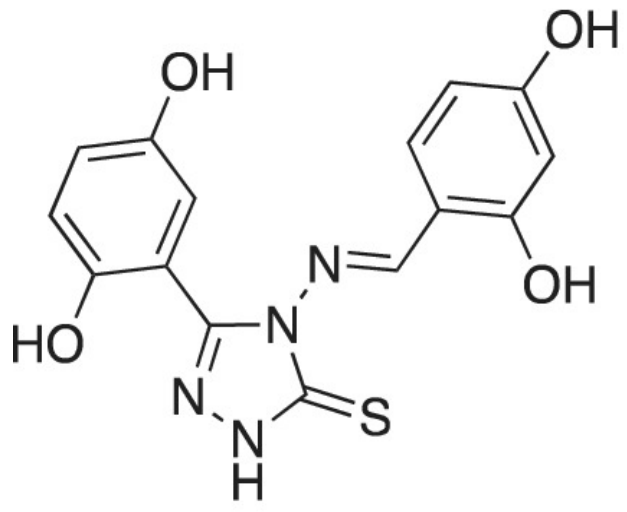

JMV4684 (III)

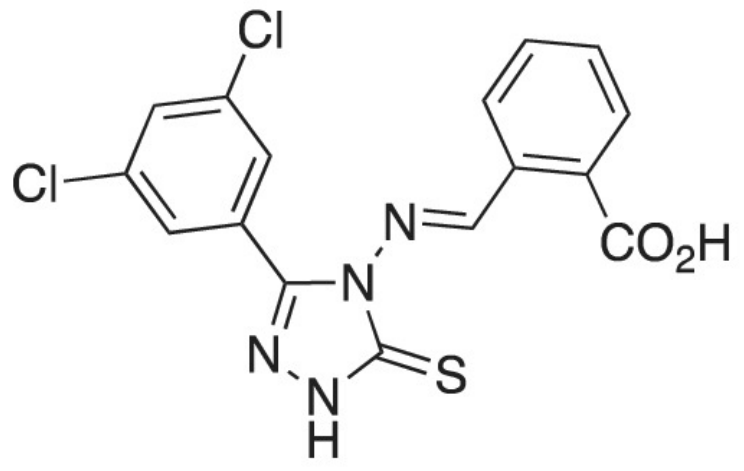

JMV5064 (IV)

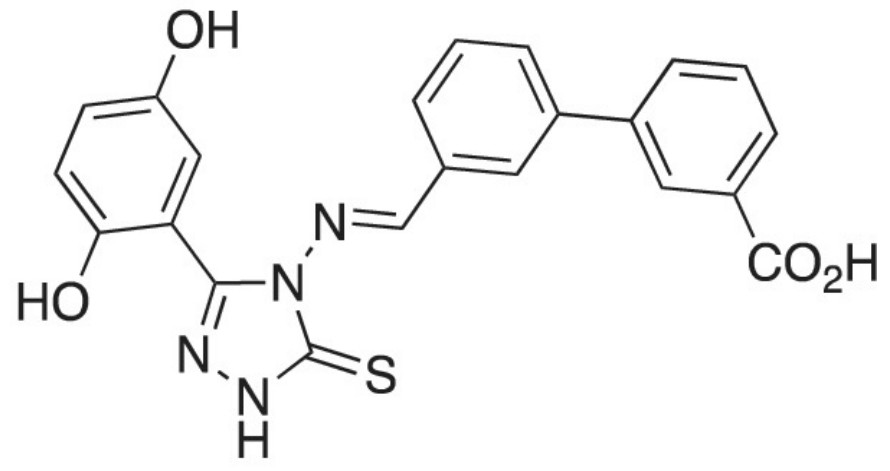

JMV5069 (V)

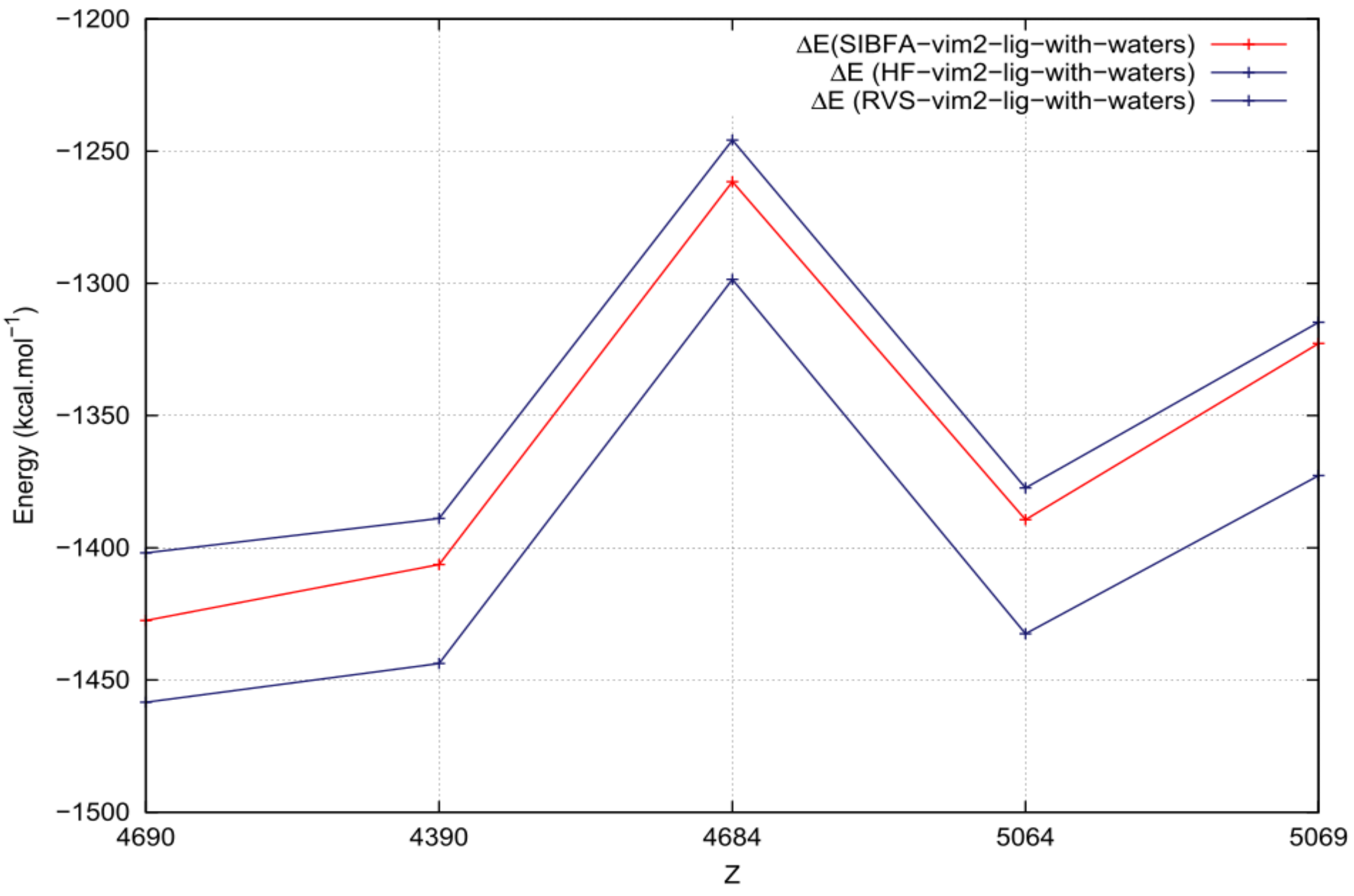

**Figure 19.** Compared evolutions of ΔE(QC/HF) and ΔE(SIBFA) in an extended recognition site of VIM-2: the inhibitors bound in the presence of structural waters.

**References.**